# 1-Adamantanamine implementation in surface engineering of biomimetic PVDF-based membranes for enhanced membrane distillation

Samer Al-Gharabli[1,2*], Nafisah Al-Rifai[1], Simona Jurevičiūte[3], Aivaras Kareiva[3], Artur P. Terzyk[4], Emil Korczeniewski[4], Ewa Olewnik-Kruszkowska[5], Zuzanna Flanc[5], Waldemar Jankowski[5], Wojciech Kujawski[5], Joanna Kujawa[5*]

[1]Pharmaceutical and Chemical Engineering Department, German Jordanian University, Amman 11180, Jordan
[2]College of Integrative Studies, Abdullah Al Salem University (AASU), Block 3, Khaldiya, Kuwait
[3]Institute of Chemistry, Vilnius University, 24 Naugarduko Street, LT-03225 Vilnius, Lithuania
[4]Faculty of Chemistry, Physicochemistry of Carbon Materials Research Group, Nicolaus Copernicus University in Torun, Gagarin Street 7, Toruń, 87-100, Poland
[5]Faculty of Chemistry, Nicolaus Copernicus University in Toruń, 7 Gagarina Street, Toruń, Poland
Corresponding authors: Samer Al-Gharabli - samer.algharabli@aasu.edu.kw
Joanna Kujawa -joanna.kujawa@umk.pl

**Abstract**

Membrane distillation (MD) stands at the forefront of desalination technology, harnessing the power of phase change to separate water vapor from saline using minimal energy resources efficiently. In response to this challenge, membranes with tuned pores morphology and surface chemistry with biomimetic 3D pine-like structures with improved affinity to water (desalination) and/or hazardous VOC (VOC removal) were developed and studied systematically. By implementing VIPS-PVDF membranes and a green modifier of 1-adamantanamine for the first time, membranes with a revolutionary network architecture were generated. The modifier was introduced either physically to the polymeric matrix or chemically through covalent attachment onto the surface and inside the porous structure. As a result, membranes that defy wetting under extreme hydrostatic pressures (>11.5 bar) were produced while preserving unparalleled vapor transport efficiency. The 1-adamantanamine promotes transport and enhances the affinity to the VOC, ensuring excellent membrane performance at different applications of the MD process. Transport was enhanced more than 3.6 times and separation factor beta changed from 3.48 to 15.22 for MTBE removal and from 2.0 to 3.46 for EtOH removal when comparing pristine PVDF with membrane chemically modified with 1-adamantanamine (PVDF_Ch02). The process separation index during the MTBE removal changed from 20 kg $m^{-2}$ $h^{-1}$ (PVDF) to 297 kg $m^{-2}$ $h^{-1}$ (PVDF_Ch02). All materials were highly stable and durable during the MD applications. This innovative approach not only revolutionizes desalination but also holds immense promise for diverse applications beyond, particularly in the realm of wastewater treatment. A study of the icing process on a cold plate with new membranes provided deeper insight into the icing mechanism and the role of membrane LEP in it.

## Graphical abstract

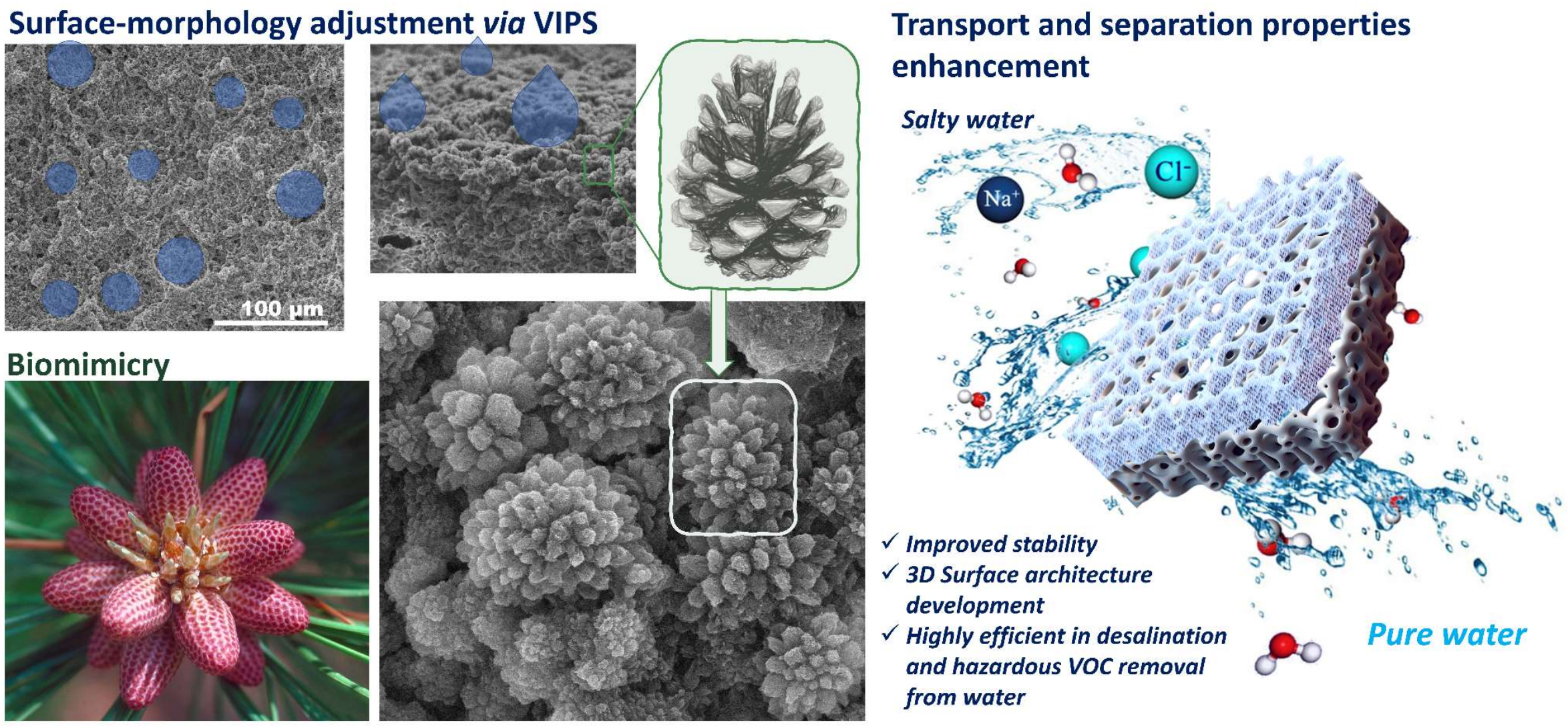

## 1. Introduction

The intersection of global warming and industrialization has catalyzed an urgent need for sustainable sources of clean water and energy [1]. Membrane technologies, particularly those utilizing highly hydrophobic, superhydrophobic, and self-cleaning membranes, are recognized as a sustainable and eco-friendly solution to tackle water and energy scarcity issues [1]. These membranes offer several benefits over other approaches, including flexibility, ease of use, and adaptability [2, 3]. The performance of membranes in membrane distillation (MD) is heavily influenced by their surface characteristics, particularly their wettability. These unique membranes play a crucial role in enhancing vapor flux and successfully combating issues such as wetting, scaling, and fouling [4]. When used in the desalination process, hydrophobic membranes come into contact with heated saline water (typically between 50 to 80°C) on the feed side, making them an indispensable component in this transformative and sustainable technology [5, 6]. The temperature difference at the membrane prompts rapid evaporation at the interface between the membrane and the saline water. The vapor then diffuses through the membrane's pores and condenses on the cooler side. The hydrophobic nature of the membrane prevents saline water from passing across, allowing only vapor to be transported [7-9]. This process effectively separates the volatile components (water) from the non-volatile ones (salts) in the heated saline solution. As a phase-change-based desalination method, MD is particularly effective for desalinating or concentrating brines with salinity levels exceeding the limits of reverse osmosis (~80 g $kg^{-1}$) [10, 11]. Owing to its theoretical 100% rejection of non-volatile substances and reduced sensitivity to feed concentration, MD is deliberated a capable way for achieving zero liquid discharge when compared with crystallization technology [12-14]. MD desalination units have a straightforward design that does not require costly materials or high-pressure equipment, making it an accessible solution for providing drinking water in rural and remote areas [15-18]. Another important application of the MD process is the removal of volatile and hazardous organic compounds from water [19-21].

Independently from the MD applications, the core of the process is a membrane that needs to be very often modified to enhance the properties, resistance, and lifespan. To tune mentioned features, the membrane can be modified *via* different pathways, including coating [3, 22-24], grafting [25-27], or incorporation of fillers, *e.g.,* perfluorooctanoic acid-modified melamine (PFOM) [28], microfibrillated cellulose [29], ZIF [30], reduced graphene oxide functionalized with ionic liquids [31], and bentonite [32]. Although surface coating is considered the simplest and most efficient process, its physical type of modification is required to repeat the process periodically. Therefore, more attention is paid to the chemical functionalization of the materials through stable covalent bonding, giving a very resistant and robust surface. Membrane functionalization can also often impair the permeability of membranes by growing their hydrodynamic resistance. Another essential issue is a sustainability perspective. Most modifiers applied in membrane modification are synthetic chemicals potentially harmful to the environment and humans [33, 34]. This is why researchers focus on green alternatives in membrane adjustment, like naturally occurring modified poly (catechol/polyamine) [35], epigallocatechin gallate/Ag nanoparticles [36], xanthan gum [37], and pyrogallol/taurine [33], cinnamic acid [38], functionalized carbon nanotubes [39]. The solutions brought to membrane science in the frame of green chemistry and sustainability cover also the application of non-fluorinated modifiers, *e.g.,* fluorine-free silane and carbon-based materials [40, 41]. Therefore, there is a severe need for an effective, economical, and eco-friendly method to functionalize membranes for MD. Thus, taking the results of the cited-above studies, the major current directions in MD science are not only increasing membrane hydrophobicity defying wetting at high pressures and enhancing vapor transport efficiency, but searching for new eco and people – and people-friendly modifiers assuring that. This is why we focus our attention on 1 – Adamantanamine. This molecule is gaining more and more attention in the scientific world, including medicine [42], chemistry [43-45], and engineering and renewable energy sectors [46, 47]. The enantioselective Strecker reaction of phosphinoyl imines serves as a catalyst [48]. It is also a starting agent for synthesizing adamantane derivatives, such as amantadine hydrochloride, which has a vast significance in the pharmaceutical industry. This component is broadly utilized for treating influenza [49, 50], as an antiviral [42, 51] and an antiparkinsonian medication. The antiviral medication amantadine provides moderate improvement in motor function for patients with Parkinson's disease [52, 53]. Memantine has beneficial effects on memory in patients with advanced Alzheimer's disease who are already receiving acetylcholinesterase inhibitors [54]. 1-Adamantanamine also found a lot of applications in material science and engineering. Liu at el. [55] applied the layer of 1-adamantanamine onto the upper surface of the perovskite film to heal the defects of the perovskite surface improving the stability and efficiency of perovskite solar cells. Singh and co-workers [56] applied 1-adamantanamine-based triazole-appended organosilanes for determination of $Sn^{2+}$ in environmental soil samples.

Considering the interesting features, the lack of toxicity confirmed by a great utility in medicine, and the availability of the amine moiety, 1-adamantanamine can be easily and effectively used in membrane functionalization *via* chemical treatment. For that reason, we proposed, for the first time, the implementation of this molecule in tuning the surface architecture (chemical pathway) and material properties (physical pathway) of the entire material. This interesting modifier allows us to tune surface architecture, morphology, and physicochemical properties precisely.

## 2. Experimental section

### *Materials*

Polyvinylidene fluoride (PVDF) Kynar® 761 powder was kindly provided by Arkema S.A. France. α-bromonaphthalene, acetone, dichloromethane (DCM), dimethylformamide (DMF), glycerol, hydrogen peroxide, methanol (MeOH), ethanol (EtOH), methyl-*tert*-butyl ether (MTBE), sulfuric acid were bought from Avantor Performance Materials Poland S.A. (Poland). 1-Adamantanamine (CAS#: 768-94-5) and 3-isocyanatopropyltrimethoxysilane (CAS#: 15396–00-6) were bought from ABCR Chemicals (Germany).

***Membrane morphology and wettability***

The structural features of the materials were analyzed by applying a scanning electron microscope (SEM), atomic force microscopy (AFM), and the modified bubble point method. The SEM provided insights into their morphological characteristics while AFM allowed to determine roughness, and based on the modified bubble point method, the pore size and pore size distribution were analyzed. Additionally, the porosity (Eq. S1) by the gravimetric method was analyzed. Instrumental details are provided in Section S1 (Supporting Materials).

The chemical features were studied with the following techniques: attenuated total reflectance Fourier-transform infrared spectroscopy (ATR-FTIR), Raman spectroscopy, and X-ray diffraction (XRD). XRD and ATR-FTIR examined the crystalline structure and level of PVDF-phases (Eq. S2, S3), while ATR-FTIR and Raman spectroscopy evaluated the functional groups and chemical makeup. Based on TGA and NMR, the grafting efficacy was determined. Additionally, surface charge was analyzed in the broad pH range between 2.5 and 11.0. Instrumental details are provided in Section S2 (Supporting Materials).

To study wettability Theta Flex, Biolin Scientific (Sweden) goniometer setup was used. The apparent water contact angle (WCA), surface free energy (SFE) with its polar and dispersive components, and critical surface tension ($\gamma_{cr}$) were determined. Moreover, the liquid entry pressure (LEP) was calculated (Eq. S6). Instrumental details are provided in Section S3 (Supporting Materials).

***Mechanical features***

The mechanical features of the samples were comprehensively analyzed using the Shimadzu EZ-SX 100 N testing device. Minimum 5 samples from each membrane were precisely-cut applying a high-quality hand press HK 80030045 (Berg & Schmid, Germany) and underwent testing. Before conducting the strength tests, the sample thickness was precisely measured to an accuracy of 0.001 mm using a high-quality Sylvac S229 thickness gauge (Switzerland). The collected results enabled to establish crucial factors accurately, *i.e.*, tensile strength at break (σ), elongation at break (ε), and Young's modulus (E) [57].

***Anti-icing properties***

To study anti-icing features, a Peltier module, thermostat, and deionized water were applied in the experiment described previously [57]. The membrane samples adhered to the surface of the cooling stage using thermally conductive double-sided copper SEM tape (Micro to Nano BV, Haarlem, Netherlands) to eliminate any thermal gap. The temperature of the droplet resting onto the membrane substrate was analyzed in the course of the freezing process using the-CTlaser LT: -50-975°C/8-14μm/ 75:1/ 120ms/ 1.9mm@150mm (OPTCTLLTCF2) pyrometer (Optris GmbH, Berlin, Germany). Before each test, the water temperature in the thermostat was adjusted to match the ambient temperature (20±1°C; ambient relative humidity equaled to 36-47%). The temperature drop profiles on the surface of the copper strips beneath the membrane samples were recorded and are presented in Supporting Information (Fig. S1).

***Membrane formation and modification***

In the presented research, membranes were generated with a vapor-induced phase separation (VIPS). The details are given in Section S4 (Supporting Materials).

The modification *via* physical blending was performed by the introduction of the 1-adamantanamine to the polymeric dope to generate 2 wt.% or 15 wt.%. The amount of the additive was calculated in reference to the mass of the pure polymer. The modification took 24 h of mixing on the magnetic stirrer (500 rpm) at RT. Afterward, the modified polymeric dope was used directly to cast the membrane by a VIPS method (details in Section S4). The membranes were labeled PVDF_P2 and PVDF_P15, accordingly, depending on the 1-adamantanamine concentration.

To accomplish the chemical way of membrane modification, prior to the process, PVDF membranes were activated to generate reactive hydroxyl groups that are available for further covalent reactions with 1-adamantanamine. The activation process was done according to the experimental protocol established in our group [58-60]. Generally, the PVDF samples (47 mm diameter) were pre-wet with solvent - methanol to guarantee contact with the activator and break the hydrophobic forces of the material before placing in the activation (15 ml for each membrane sample) in closed Schott glass bottles. The activator, *i.e.* piranha solution 20 wt.% aqueous solution of $H_2SO_4$ and $H_2O_2$ mixture in 3:1 ratio was freshly prepared before usage. The activation step took 30 min and was accomplished at 25 ºC. Then, samples were cleaned with DI water (15 MΩ·cm) and methanol and subsequently dried for 24 h in an oven at 70 °C. The PVDF samples furnished with OH - hydroxyl groups was used for silanization and, finally, functionalization with 1-adamantanamine.

Silane-based modifier (linker), *i.e.*, 3-isocyanatopropyltrimethoxysilane 0.05M solution, was prepared in dichloromethane at RT. The activated membrane, with OH groups, was placed in closed Schott glass reagent bottles with 50 ml of 3-isocyanatopropyltrimethoxysilane solution and then mixed for 72 h. Later, the membrane was cleaned with methanol, dichloromethane, and water and dried for 24 h in an oven at 70 °C. The membranes were equipped with silane molecules with available isocyanate-reactive groups, which FTIR-ATR confirmed. Those groups were used in the final step to generate covalent bonding with the amino group from the 1-adamantanamine (Fig. 1 – modification reaction). Finally, the silanized membranes were placed in the 0.1M or 0.2M of 1-adamantanamine solution prepared in dichloromethane. The modification took 24 h under constant mixing and ambient atmosphere of argon. Next, the functionalized membranes were cleaned with methanol, dichloromethane, and water, and dried for 24 h in an oven at 70 °C. Membranes were labeled as PVDF_Ch01 or PVDF-Ch02, depending on the 1-adamantanamine concentration.

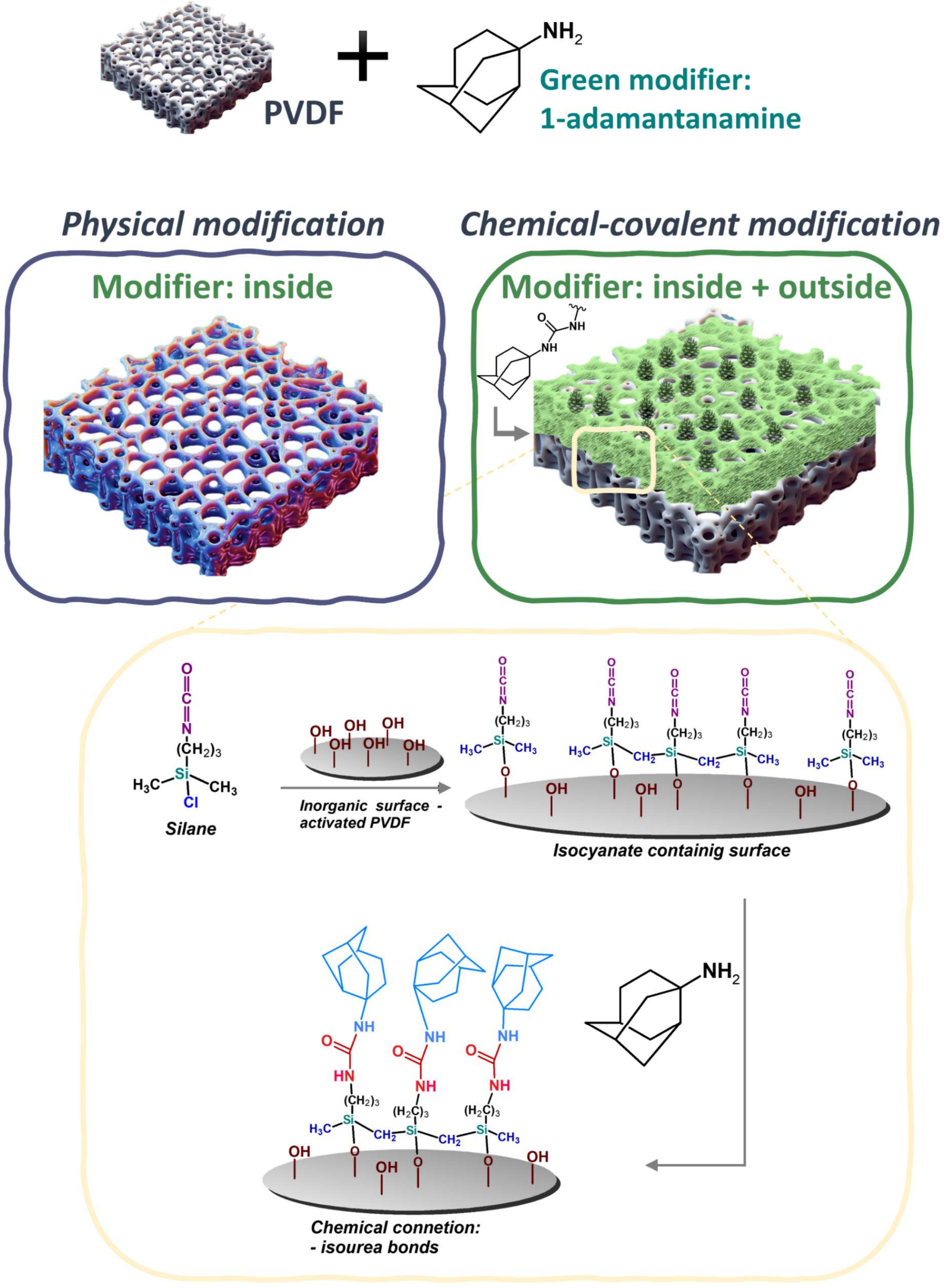

**Fig. 1.** The way of membrane modification.

## 3. Membranes applications

The efficacy of the membranes with 14 $cm^2$ of active area was assessed in the air-gap membrane distillation process. Two applications were selected: (i) the desalination process and (ii) the removal of hazardous volatile organic compounds (VOCs) (EtOH and MTBE) from water.

(i) Before conducting the desalination test with a 0.5M NaCl solution, water transport was analyzed under various driving forces between 60 and 460 mbar ± 0.5 mbar. These driving forces were achieved by putting on different temperatures to the feed side (40 - 80 °C ± 1.0 °C), while maintaining a constant temperature on the permeate side at 7 °C ± 0.5 °C. Each run lasted for at least 9 hrs. After collecting data at the different temperatures, the apparent activation energy ($E_{ap}$) (Eq. S10) was determined based on the Arrhenius equation [61, 62]. Then, the desalination process was performed with the 0.5 M NaCl solution. Conductivity measurements were used to determine the NaCl rejection coefficient (Eq. (1)) with Elmetron CPC-505 (Poland).

$$R_{NaCl} = \left(1 - \frac{C_p}{C_f}\right) \cdot 100\% \quad (1)$$

where $C_p$ and $C_f$ are salt concentrations in the permeate and the feed, respectively

(ii) In the second step, the separation properties were determined in the VOC removal measurements. The tests were accomplished at the following temperature conditions: the feed side equal to 60 ± 1.5 °C and the permeate side 8 ± 1.5 °C generate a driving force of ca. 200 mbar ± 0.68 mbar. The process efficiency was monitored with gas chromatography. Instrumental details are presented in Section S5 (Supporting Materials).

## 4. Assessment of bioaccumulation potentials

The bioaccumulation potential of silane-based linkers and modifiers used during the functionalization process was quantitatively analyzed with the implementation of the following parameters: (i) bioconcentration factor (BCF), (ii) octanol-water partition coefficient ($K_{ow}$), and (iii) bioaccumulation parameter (BAF). The detailed characterization are presented in Section S6 (Supporting Materials).

## 5. Results and discussion

### 5.1. Membrane characterization – morphology

The selection of different ways of membrane modification allows to adjust precisely the physicochemical (wettability - water contact angle from 131.7° to 164.2°) and morphological features (roughness from 417 nm to 756 nm; average pore size in the range of 0.132 μm and 0.226 μm) (Fig. 2). While the VIPS procedure, like other phase inversion methods, can achieve various morphologies like porous, dense, symmetric, or asymmetric ones, its distinctive capability to regulate the phase inversion rate effectively and membrane morphology offers a superior opportunity for precise customization compared to alternative phase inversion techniques [63].

Membranes were characterized by a spheroid, symmetric nodular structure typical for VIPS membranes [64]. Nodular morphology, also referred to as granular or spherulitic structure, is a prevalent form observed in membranes produced through VIPS. Typically composed of semi-crystalline and crystalline polymers, this structure is consistently present across the entire cross-section of the generated membrane material. The slower intake rate of the non-solvent (water vapor) in VIPS promotes polymer crystallization over liquid-liquid demixing (L-L) [65-67]. Furthermore, this process sustains the cast film within the crystallization region without being influenced by L-L phase separation. Consequently, this phenomenon results in the development of a granular structure and solid sphere, which extends throughout the overall cross-section of the material (Fig. 2, S2). Formation of such morphology is often associated with the gelation induced by crystallization [68] or when the critical dissolution temperature ($T_{cri}$) is exceeded [69]. The demixing process of the homogeneous polymer solution significantly impacts the morphology of the final material depending on the thermodynamic and kinetic factors [70, 71]. Nevertheless, the spheres forming a porous matrix differ depending on the modification process. The pristine membrane possesses in the bulk, spherical microstructures with an average size of 2.28 ± 0.27 μm. Moreover, each spheres contain nanostructures with a diameter *ca.* 0.13 ± 0.02 μm (Fig. S2). Owing to the modification, the size of the sparoids changed (Fig. S3-S5), which was particularly visible in the chemical treatment (Fig. S4, S5). The spheres in the nodular structure were characterized by the sizing of 2.60 ± 0.45 μm and 2.34 ± 0.24 μm for PVDF_Ch01 (Fig. S4) and PVDF_Ch02 (Fig. S5), respectively.

In the case of membranes modified *via* a physical pathway, the nodular morphology with a bi-continues-like structure was generated. The reason for such a structure was the presence of a 1-adamantanamine modifier, altering the solubility parameters. The affinity between the solvent and non-solvent is the most essential parameter, and the viscosity of the polymer solution affects the membrane morphology in the VIPS method [72, 73]. This morphology is characterized by a slightly more open pore structure (Table 2) that can be beneficial during the separation process owing to the reduction in permeation resistance. The reduction in the pore size for the physical treatment was equaled to 12.3% (PVDF_P2) and 41.5% (PVDF_P15), respectively. On the other hand, the chemical reduces more significantly by 32.7% (PVDF_Ch01) and 46.4% (PVDF_Ch02).

The mentioned changes also impact the porosity of the material (Eq. S1). With the reduction of the pore size, the porosity was diminished for all the investigated samples (Table 2), which was related to the presence of 1- adamantanamine in the entire volume of the matrix. 1-adamantanamine enhances PVDF crystallization, reduces phase separation during the VIPS method, and finally, the porosity decreases due to a denser microstructure (Fig. 2).

The differences in the bulk morphology will directly impact the transport properties of the membrane during the separation process. Nevertheless, substantial differences were visible on the membrane surface (Fig. 2). These properties will have a viral effect on the wettability as well as other physicochemical and material features.

Owing to the well-developed surface, typically available for VIPS membranes, their hydrophobic character was revealed for all samples. The highly and superhydrophobic character was an aftermath of membrane modification. The samples being physically modified were characterized by a slightly reduced water contact angle (PVDF_P2 = 141.9 ± 0.6°; PVDF_P15 = 131.7 ± 0.7°) compared to the pristine sample (PVDF = 143.2 ± 0.7°) that was related to the presence of the hydrophilic filler in the material (Table 2). Furthermore, the chemically treated materials possessed higher values of WCA equal to 148.9 ± 0.5° (PVDF_Ch01) and 164.2 ± 0.8° (PVDF_Ch02). Such a high WCA was, however, related to the substantially improved roughness from 417 ± 71 nm for pristine PVDF to 693 ± 66 nm for PVDF_Ch01 and 756 ± 80 nm for PVDF_Ch02, respectively (Table 1 and Fig. 3). The modification process also impacted the internal morphology of the membrane by reducing the pore sizes (Fig. S10). It was additional evidence of the surface treatment. The most significant pore reduction was observed for the PVDF_P15 and PVDF_Ch02. In the PVDF-P15 sample, the reduction (41.5%) was attributed to the higher filler content inside the porous structure. However, for the latter one, the formed 3D structure layer from the 1-adamantanamine (Fig. 2 E3, Fig. S6) was behind the pore reduction (46.4%), Table 2.

**Table 1**. Membrane morphological, material and physicochemical features.

| Membrane | IEP | Surface free energy, SFE [mN m$^{-1}$] | | | $\gamma_{cr}$ [mN m$^{-1}$] | LEP [bar] | WCA [deg] | Roughness parameter – Rq [nm] |
|---|---|---|---|---|---|---|---|---|
| | | $\gamma_{total}$ [mN m$^{-1}$] | $\gamma_d$ [mN/m] | $\gamma_p$ [mN m$^{-1}$] | | | | |
| **PVDF** | 3.63 | 41.06 | 33.59 | 7.47 | 42.1 | 3.82 | 143.2 ± 0.7 | 417 ± 71 |
| **PVDF_P2** | 3.48 | 41.67 | 33.60 | 8.06 | 41.7 | 5.42 | 141.9 ± 0.6 | 574 ± 68 |
| **PVDF_P15** | 3.80 | 46.17 | 38.61 | 7.56 | 43.3 | 4.72 | 131.7 ± 0.7 | 539 ± 80 |
| **PVDF_Ch01** | 10.36 | 67.94 | 63.58 | 4.37 | 47.3 | 8.11 | 148.9 ± 0.5 | 693 ± 66 |
| **PVDF_Ch02** | 10.49 | 59.34 | 56.90 | 2.45 | 48.1 | 11.45 | 164.2 ± 0.8 | 756 ± 80 |

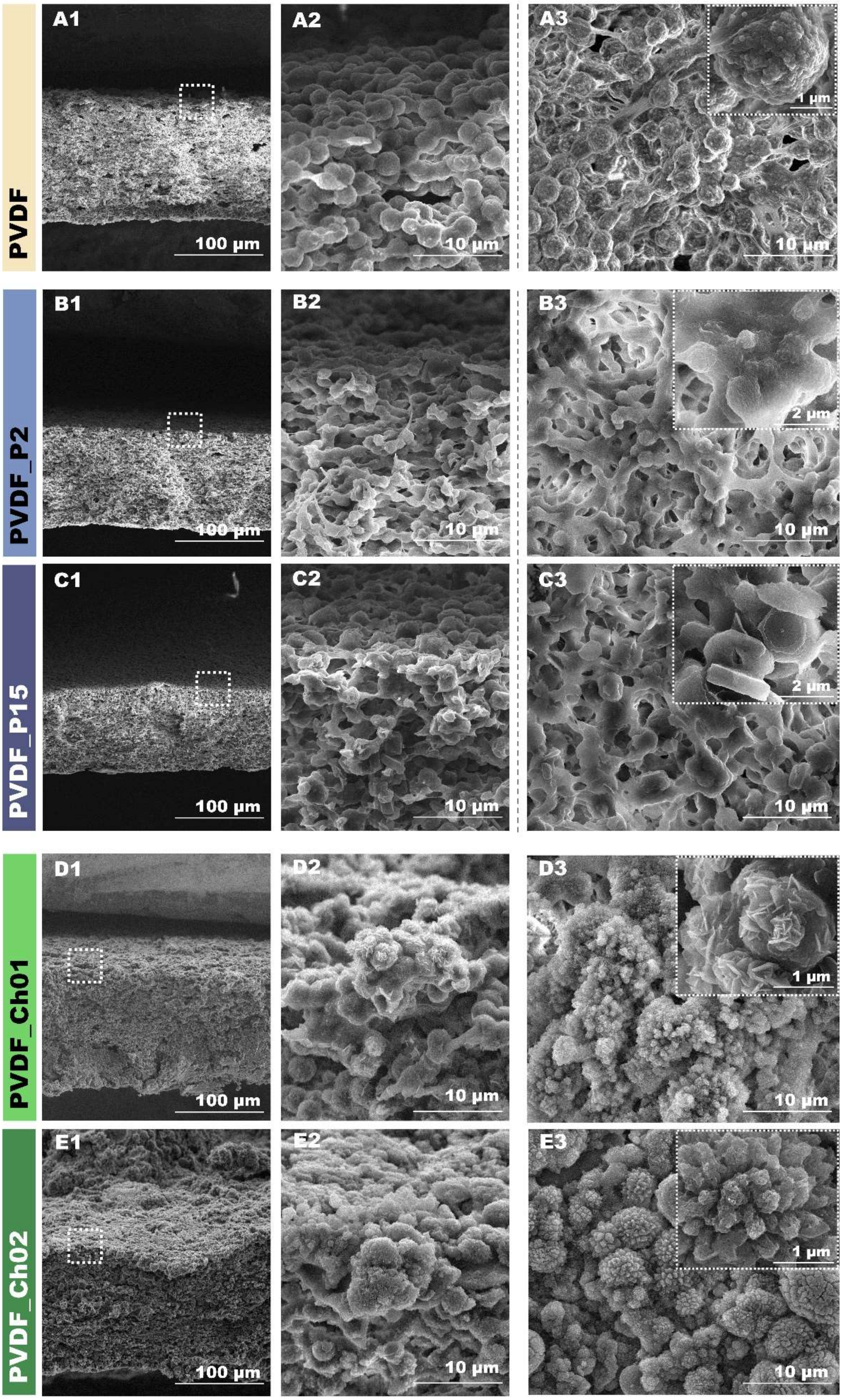

**Fig. 2.** SEM images of pristine (A) and modified membranes via physical (B,C), and chemical (D,E) way. X1 and X2 – cross section, X3 – surface.

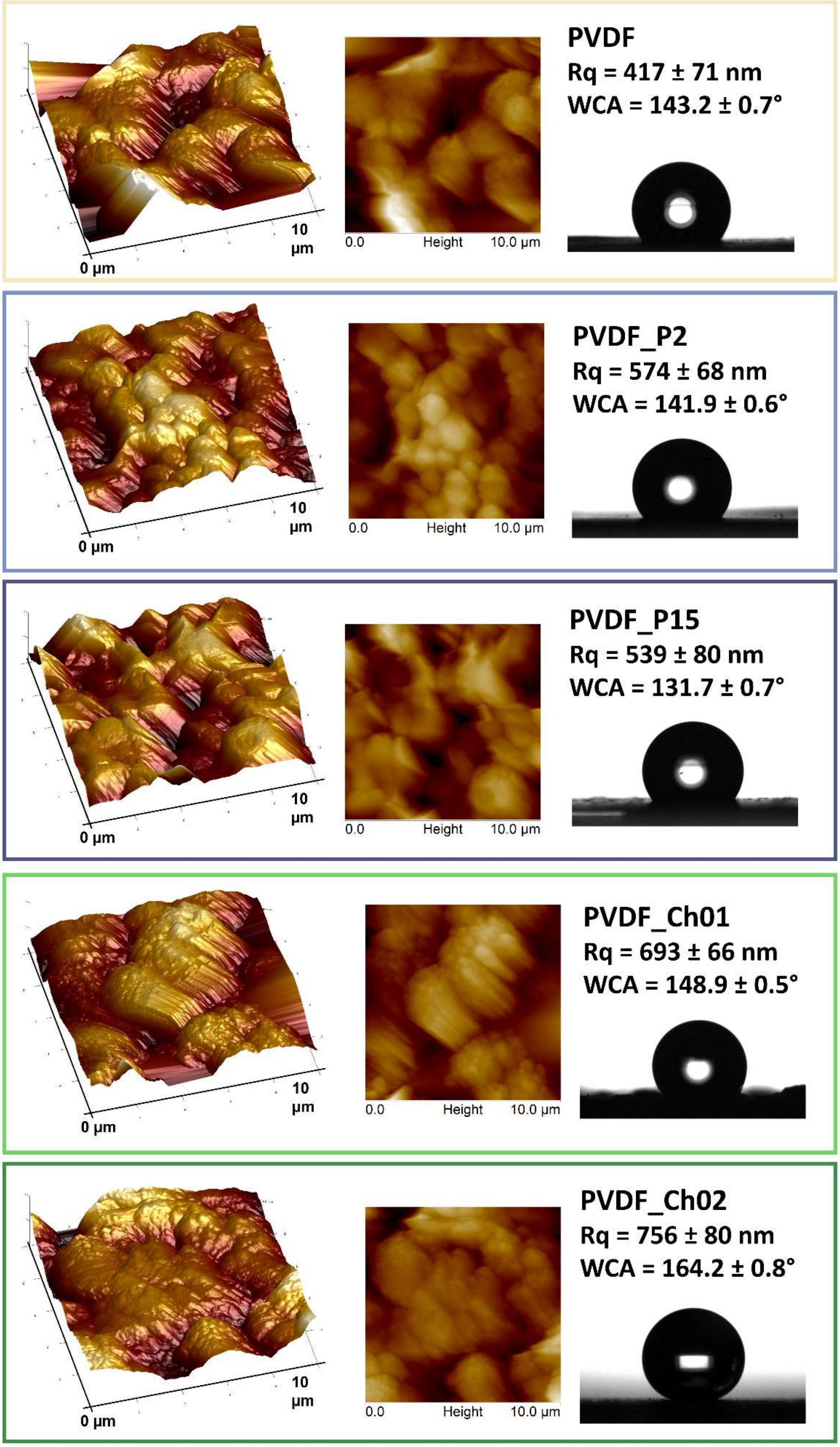

**Fig. 3.** AFM of the studied membranes (scanning area 10 x 10 μm) and water contact angle.

**Table 2**. Pore size analysis of the investigated membranes formed with VIPS method.

| Membrane | Min pore size [µm] | Max pore size [µm] | Av. pore size [µm] | Pore size reduction referred to pristine [%] | Porosity [%] |
|---|---|---|---|---|---|
| | | | **Pristine membranes** | | |
| **PVDF** | 0.075 | 0.301 | 0.226 ± 0.032 | - | 62.3 ± 1.2 |
| | | | **Modified physically** | | |
| **PVDF_P2** | 0.121 | 0.209 | 0.198 ± 0.025 | 12.3 | 56.5 ± 1.0 |
| **PVDF_P15** | 0.083 | 0.203 | 0.132 ± 0.022 | 41.5 | 48.6 ± 1.0 |
| | | | **Modified chemically** | | |
| **PVDF_Ch01** | 0.090 | 0.198 | 0.152 ± 0.034 | 32.7 | 52.4 ± 1.1 |
| **PVDF_Ch02** | 0.075 | 0.209 | 0.121 ± 0.018 | 46.4 | 45.4 ± 0.8 |

### 5.2 Polymer-modifier affinity

As stated in the previous section, the solubility parameters play a crucial role in tuning the morphology. The Hansen Solubility Parameters (HSP) approach has been implemented to study the observed changes. During the calculation (Eq. S7-S9), the distance in Hansen's sphere ($\Delta$) (Eq. S7) and Relative Energy Difference (RED) (Eq. S9) were determined. Our previous articles [57, 74] provide a detailed explanation of the mentioned parameters.

The distance in the Hansen's sphere increased after the addition of 1-adamantanamine modifier (Table S2-S5), showing a slightly lower affinity between the materials containing 1-adamantanamine and DMF solvent (Table S2). The distance for pristine PVDF was equal to 1.98 $MPa^{0.5}$. Modified membranes possess the following values of $\Delta$, 2.17 $MPa^{0.5}$ (PVDF_P2), 3.34 $MPa^{0.5}$ (PVDF_P15), 4.90 $MPa^{0.5}$ (PVDF_Ch01), and 5.87 $MPa^{0.5}$ (PVDF_Ch02) (Table S3). Nevertheless, all values of the RED were in the range of 0.25 and 0.73, which confirmed the good solubility (Table S3, S5). The RED equal to 1 is the boarder of solubility. After that, the material will either swell or not react at all with the solvent. In Hansen's theory, such a solvent, with RED > 1, is classified as a "bad solvent" [75]. Such interaction had an influence on the morphology. Specifically, the better affinity and smaller distance Δ allow the generation of a morphology with bigger spheroids (size 2.28 ± 0.27 μm) (Fig. S2, Table S6). The addition of hydrophilic (hydrophilicity 2.4) 1-adamantanamine to the hydrophobic polymeric matrix of PVDF reduces the affinity. Such character of the polymeric dope allows the generation of morphologies with smaller spheroidal structures (Fig. S3, S4). Finally, for the chemically modified materials, *i.e.*, PVDF_Ch01 and PVDF_Ch02, the structure with a slightly bigger structure was noticed (Fig. S5, S6). Moreover, these spheroidal structures possessed surface irregularities in the nanoscale (nodular morphology with a bi-continues-like structure).

### 5.3 Surface chemistry and charge

The functional groups and chemical makeup were determined using ATR-FTIR and Raman. In the case of physically modified materials, characteristic bands of modifiers were visible, specifically in the dactyloscopic fingerprint region (Fig. 4). Moreover, the spectrum for PVDF-P15 membrane possessed the characteristic bands with higher intensity than the spectrum for the PVDF-P2 sample, which was related to higher level of modifier in the matrix. The uniform modification and presence of the modifier in the entire volume during the physical modification were confirmed additionally by FTIR spectra of the powdered samples to the pellet form (Fig. S7). In the case of chemical treatment, the covalent attachment of the 1-adamantanamine was confirmed by the presence of new bands of isourea connection (Fig. 1, 4). The appearance of these characteristic bands in the 1300 – 1700 $cm^{-1}$ range confirmed the covalent attachment between the amino group of 1-adamantanamine and the isocyanate group of the silane-linker (Fig. 4). Additionally, the presence of the modifiers has been confirmed by the presence of bands in the range of 2550 – 2910 $cm^{-1}$ related to the different C-H, $CH_2$, and $CH_3$ stretching bands. The inherence of modifiers inside the porous matrix was confirmed by the FTIR of the powdered membrane sample (Fig. S8). Additionally, the grafting effectiveness was analyzed. The level of the available OH groups on the pristine material was equal to 1.6 ± 0.12 $10^{-3}$ mol/g. However, after the functionalization with 3-isocyanatopropyltrimethoxysilane this value reduced to 0.29 ± 0.02 $10^{-3}$ mol/g. It means that around 82% of the OH groups were used to generate connections with silane modifiers and ensure very high grafting efficiency. The value is in good accordance with the membrane functionalization with non-fluorinated silane-based modifiers [58, 76, 77]. Based on the $^{29}$Si NMR data (Fig. S9), it was found that molecules are connected in two ways. Two signals were observed: between −55 ppm and −70 ppm ($T^n$ signals) and between −87 ppm and −115 ppm (for uncondensed silanol groups). The signals can be assigned to the appropriate $T^n$ structures: $T^2$ ($O_2Si$- (OEt)R), and $T^3$ ($O_3SiR$) at chemical shifts from −53 ppm to −59 ppm, and from −62 ppm to −69 ppm, respectively. 3-isocyanatopropyltrimethoxysilane was connected to activated PVDF mostly by $T^2$ (62%) and $T^3$ (38%), respectively.

The level of the alpha and beta phases was analyzed (Eq. S2 and S3) with the implementation of FTIR spectra. The pristine PVDF material possesses the ca. 10.5% alpha phase. After the addition of a 1-adamantanamine modifier to the polymeric matrix by physical blending, the level of the alpha phase reduced to 7.6% (PVDF-P2) and 9.0% (PVDF-P15). Finally, the chemically tuned materials with abundant modifiers on the surface were characterized by the level of alpha phase equal to 46.5% (PVDF_Ch01) and 63.2% (PVDF_Ch02).

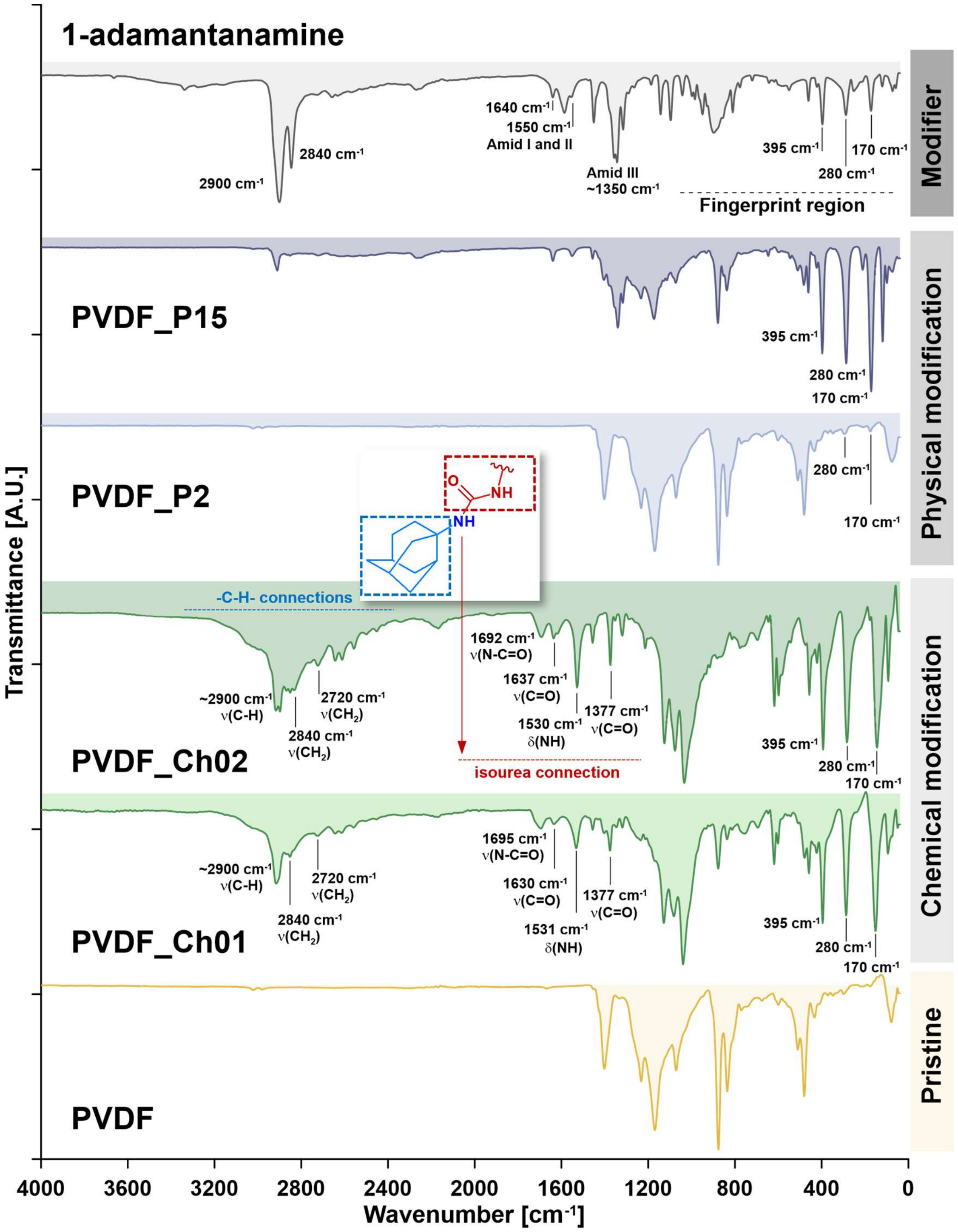

**Fig. 4.** ATR-FTIR spectra of the investigated membranes.

Complementary analysis with Raman spectroscopy registered the bands that are not active in FTIR mode (Fig. S11). The most important bands were found and highlighted, *i.e.*, ring in-plane bend (551 $cm^{-1}$) in 1-adamantanamine modifier, υCC chain vibration (640, 714 and 776 $cm^{-1}$, 1204 $cm^{-1}$), and $\delta_{CH}$ (1088 $cm^{-1}$). Furthermore, the appearance of the band at 1650 $cm^{-1}$ confirmed the connection between the reactive moieties of -NCO from the linker on the membrane (3-isocyanatopropyltrimethoxysilane) and $-NH_2$ from the modifier site (Fig. S11). The above-mentioned changes have also been confirmed by the XRD analysis. Specifically, the changes in the beta-to-alpha ratio of PVDF (Fig. S12). A substantial increase in this factor (Fig. S12) for the chemically modified membranes is attributed to the presence of these molecules on the outer surface of the membranes of PVDF-Ch01 and PVDF_Ch02 (Fig. 2, S5, S6). Due to the fully chemical treatment, the generated layer was highly stable during the treatment as well as tests during the separation processes.

Zeta potential (ζ) analysis through potential streaming is a critical parameter in material characterization studies, particularly in identifying subtle variations in surface charge post-modification. The alterations observed in the zeta potential profiles (Fig. 5A) are consistent with data obtained from other methods, such as FTIR (Fig. 2), Raman (Fig. S11), and XRD (Fig. S13). The isoelectric point (IEP) for unmodified PVDF was determined to be at pH 3.15, aligning well with the literature-reported range of pH 2.5–4.5 [78]. The IEP of PVDF, ca. pH 4.0, indicates polymers lacking dissociable groups. The hydrophobic character of the unmodified PVDF promotes the adsorption of chloride anions, which results in a negative zeta potential within the pH range of 4–10. As was expected, the value of this potential for the initial PVDF sample decreases with the pH of the solution (strongly hydrophobic surfaces adsorb negative $OH^-$ ions [79]). Similar plots are recorded for the PVDF_P2 system, since in this case the number of $NH_2$ surface groups is relatively small (causing, as it was mentioned above, only small decrease in WCA value). However, for the membrane PVDF_P15 (showing the largest hydrophilicity), zeta potentials at small pH levels are larger than for two remaining systems, and are drastically smaller at larger pH levels (note that for all three systems, similar isoelectric points, IEP, are recorded). Larger zeta potential values below IEP are caused by the adsorption of protons by basic $NH_2$ groups and protonation. Smaller zeta potentials above the IEP (large pH levels) are the result of large $R_q$, and in this way, larger $OH^-$ ions adsorption on the hydrophobic parts of a surface (Fig. 5). Completely different profiles of surface charge changes were observed for the chemically modified materials. First of all, the IEP values were significantly shifted to 10.37 and 10.50 for the PVDF_Ch01 and PVDF_Ch02, respectively. Moreover, the curves were placed in the positively charged region practically within the entire range of measurements, from 2.5 to 10.5 pH. Such behavior confirms the high efficiency of the chemical treatment and the availability of the modifier molecules (Fig. 5). Similar curves for the PVDF modified with molecules equipped with amine groups, *i.e.*, hyperbranched polymers amino end group was presented by Schulze and Prager [78], where additionally the change in the IEP from 3.5 (pristine PVDF) to 9.4 (PVDF-$NH_2$).

To analyze changes related to the surface charge more in detail, two other parameters were determined, *i.e.*, the evolution of the gap height during the measurement (Fig. 5B) and the relation between the wettability and zeta at the selected pH, *i.e.,* pH = 8 (Fig. S13). The monitoring of the gap height is a source of valuable information. Expressly, if no changes in the gap height are detected, the sample's high stability is confirmed. In the case of the presented samples, the most stable samples were those treated in a physical way with a higher level of loading, *i.e.*, PVDF_P15 and both chemically modified ones (Fig. 5). On the other hand, the pristine PVDF, specifically in the highly basic media, and PVDF_P2 practically in the entire pH range were slightly less stable. Nevertheless, the general very high stability of the membrane was proved.

Fig. S13 shows a relationship between the zeta potential at pH 8 and the water contact angle for the investigated samples. A very interesting relation has been found. Namely, after the PVDF membrane's physical treatment, the hydrophobicity level reduction was observed with the shift to a more negatively charged surface. Such behavior is related to functional moieties solely due to the very little impact on IEP. On the other hand, an increase in the hydrophobicity level and a shift to a more positively charged surface have been found for PVDF treated chemically. This needs to have resulted from the high level of amino groups from the modifier and the behavior of the basic moieties that will get protonated, and thus, a more positive charge is generated (Fig. S12).

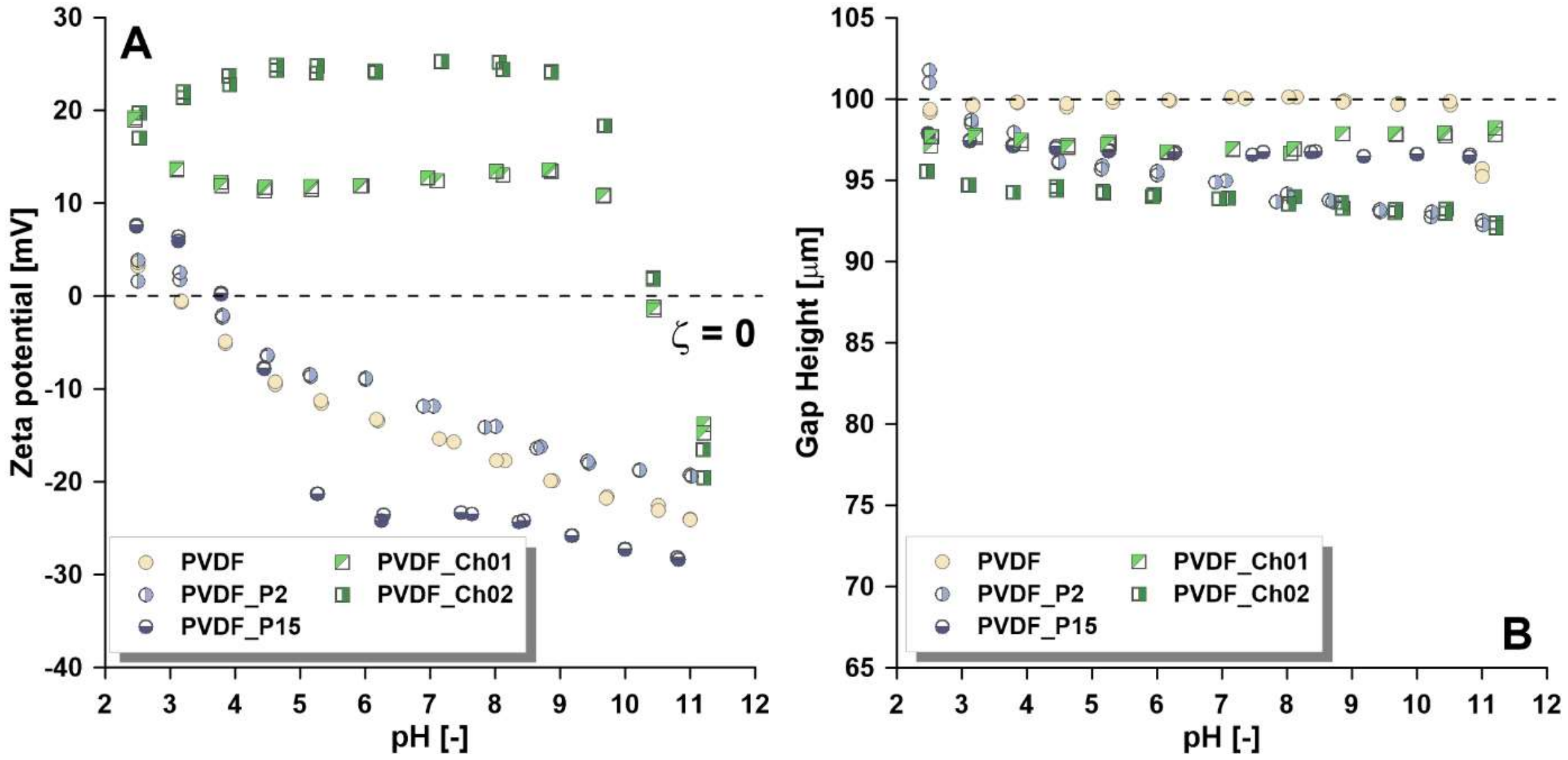

**Fig. 5** The surface charge of the investigated membranes (A) and changes in the gap height (B) during the zeta potential tests.

### 5.4 Wettability and anti-icing studies

The influence of modification by 1-adamantanamine on water contact angle shows the progressive hydrophilization after physical treatment, with the WCA for the starting PVDF membrane equal to 143.2°, for PVDF_P2 (containing 2 wt.% of 1-adamantanamine) 141.9°, and 131.7° for the PVDF_P15 (having 15 wt.% of 1-adamantanamine) (Fig. 6, Table 1). On the other hand, an increase in WCA for PVDF_Ch01 (WCA = 148.9°) and particularly PVDF_Ch02 (WCA = 164.2°) was noticed (Table 1). However, since observed WCA values are larger than 90°, all surfaces are still (as the initial sample) hydrophobic, which is essential for membrane distillation application. The data collected in Table 1 reveal the rise in $R_q$ (from 417 ± 71 nm for PVDF to 574 ± 68 nm and 539 ± 80 nm for PVDF_P2 and PVDF_P15, respectively) after modification (Fig. 6). Moreover, the $R_q$ values for PVDF_P2 and PVDF_P15 samples are (in the range of experimental error) the same. Thus, the observed decrease in WCA for physically tuned materials is caused by the Wenzel effect [80]. This effect occurs by the appearance of strongly polar $NH_2$ groups on surfaces since hydrophobic PVDF chains show stronger affinity to the nonpolar part of 1-adamantanamine molecules (this is observed especially for the PVDF_P15 membrane). This observation agrees with the dependence of observed changes in the values of zeta potential (Fig. 5A) on the pH of the solution. A different situation has been observed for PVDF_Ch01 and PVDF_Ch02 materials, for which an increase in water contact angle was noticed with the rise of roughness that is typical of Cassie-Baxter surfaces [81]. Moreover, the formation of Cassie-Baxter materials was confirmed by the increase in surface free energy (SFE) and critical surface tension ($\gamma_{cr}$), which is typical for hydrophilization of the material by tuning chemistry, here the introduction of the hydrophilic 1-adamantanamine. Nevertheless, the high resistance to wetting during the membrane application in the separation process is ensured by the increase in the LEP values from 3.82 bar (PVDF) up to 11.45 bar for PVDF_Ch02 and the reduction in the polar component of SFE (Table 1, Fig. 6). The discussed relation has been visualized in Fig. 6, which shows clearly how the chemical treatment could improve the wetting resistance even though a hydrophilic modifier was used. The reason, as stated above, is the change from the Wenzel-like structure of the physically tuned membranes to the Cassie-Baxter-like materials of the chemically adjusted PVDF-based membranes (Fig. 6). This statement is aligned with the HSP affinity between the membranes and water being >> RED = 1 confirming the high stability (Table S4, S5). The RED was in the range of 4.04 and 4.39 for all of the investigated materials.

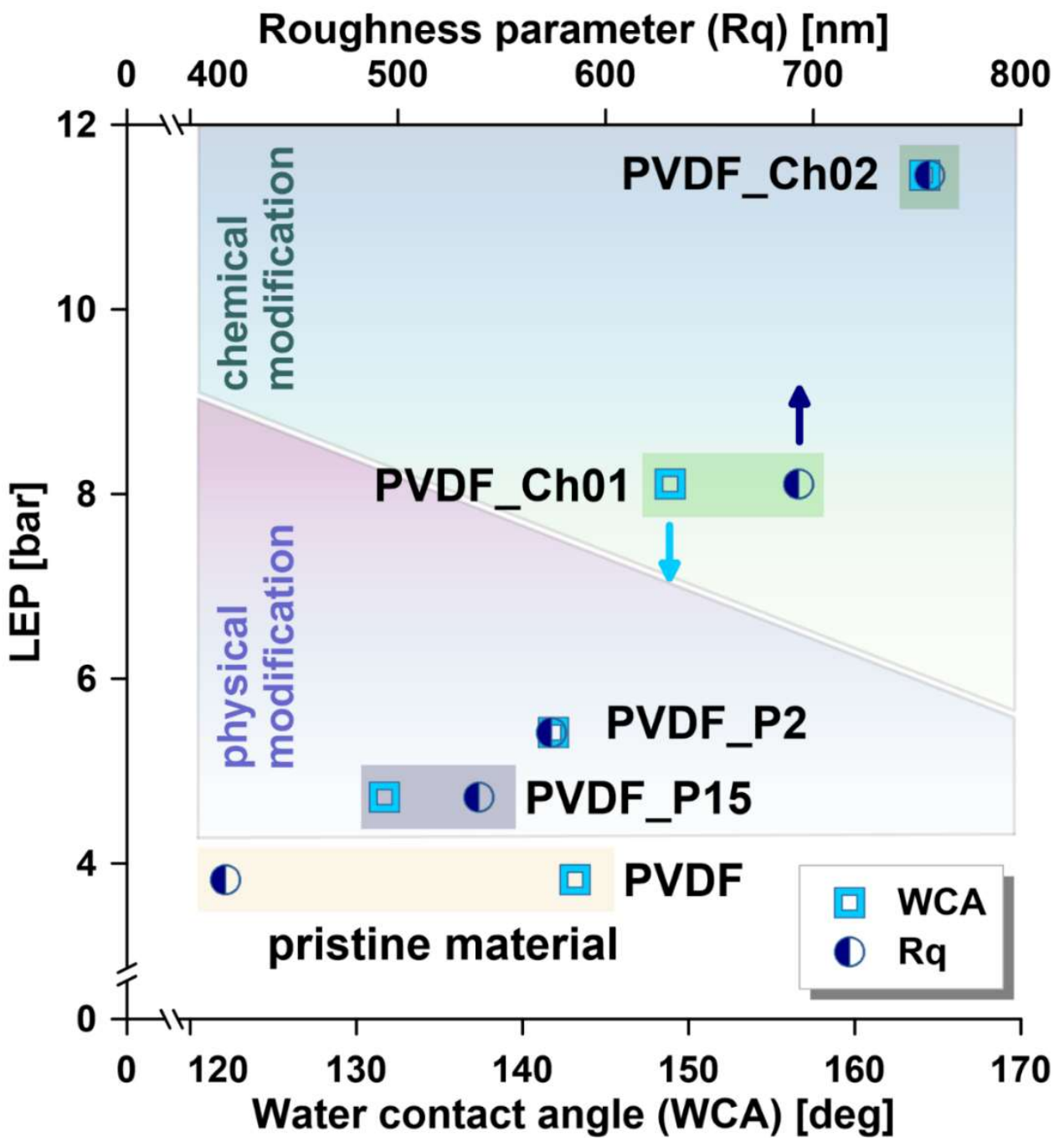

**Fig. 6.** Correlation between physicochemical properties, morphology and wetting features.

To study anti-icing features, we must consider that the cooling process is done in the bottom-top profile, therefore the physical meaning of anti-icing properties can be evaluated for the sample modified within an entire volume of membrane matrix. In the case of surface treatment only, the results will possess significant errors and lack of logic in such studies. To follow the mentioned protocol, only the pristine membrane (PVDF) and the physically treated samples (PVDF_P2 and PVDF_P15) were included in the determination of anti-icing properties.

In our previous study [57], a new relationship between LEP and time to recalescence during water droplet freezing was presented. This linear relationship is very intuitive, and for the first time, it relates WCA, pore diameter, and water surface tension with the anti-icing properties. The relationship was presented for a series of THV membranes modified with low and high-mass chitosan, and the modification led to a rise in membrane hydrophobicity (see Fig. 11 A in [57]). The current study observed the opposite situation, since adding 1-adamantanamine decreases WCA values of the physically treated samples. Thus, it is interesting to check the validity of the LEP-recalescence time relationship on the samples of this study to explain the influence of membrane hydrophilization on this relationship. The detailed inspection revealed that no linear correlation between LEP and recalescence time occurs for the studied membranes. In contrast, there is a quasi linear relationship between recalescence time and the % content of 1-adamantanamine (Fig. 7).

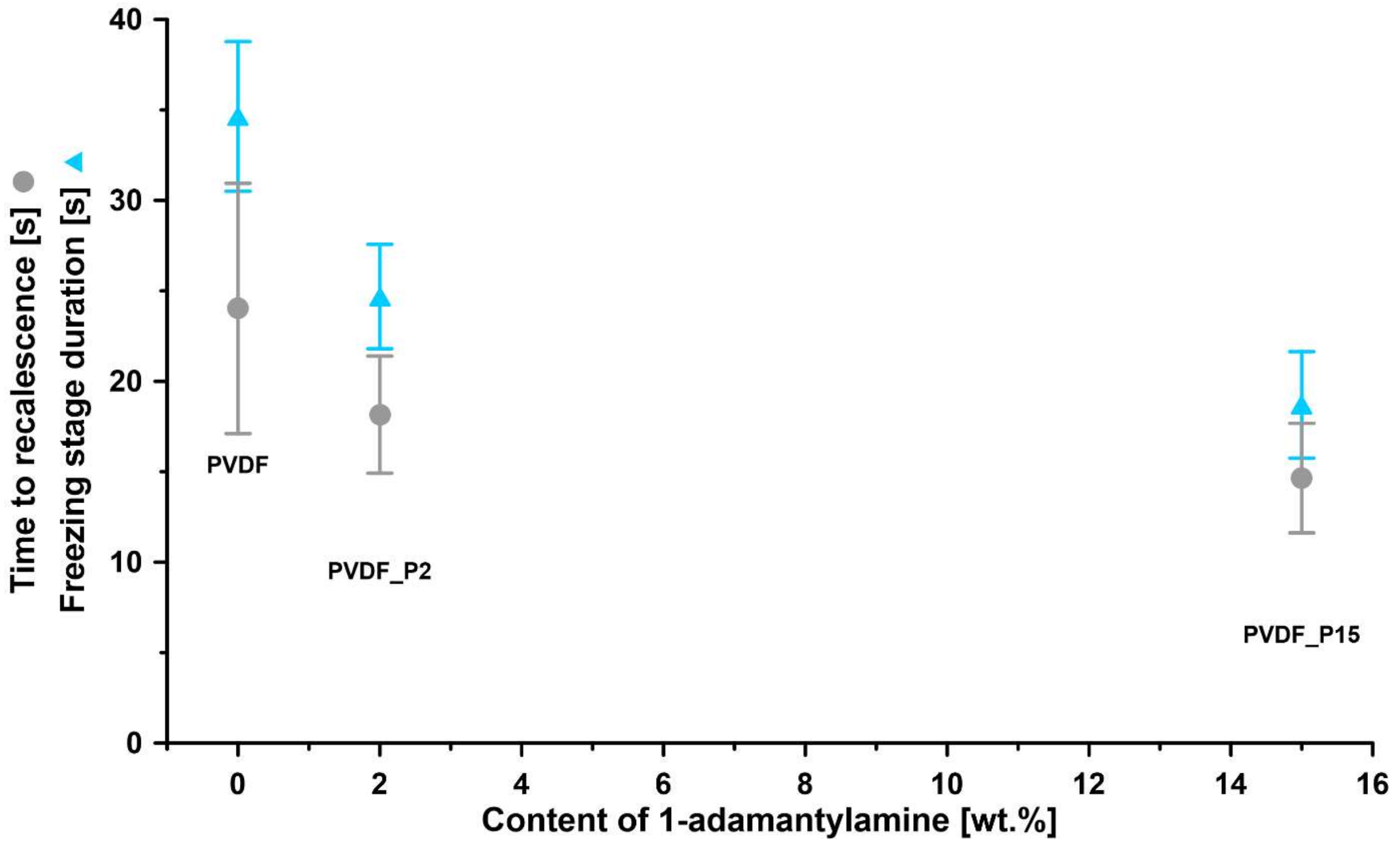

**Fig 7.** The relationship between time to recalescence, freezing stage duration, and % content of 1-adamantanamine.

It can be observed that the addition of 1-adamantanamine leads to a shortening of recalescence time as well as freezing stage duration (Fig. 7). This is caused by the creation of new centers promoting water nucleation [82] around polar and basic $NH_2$ sites *via* hydrogen bonding, and/or acid – base Lewis interactions. Interestingly, the lack of correlations with LEP shows that, in this case, the creation of strongly polar surface sites plays a more important role than the membrane porosity parameters. This is an analogous situation to water vapor adsorption on hydrophobic materials, where the porosity of a surface is not as important as the interaction of water with surface polar centers [83].

Recalescence process, due to its very short duration time, is not easy to study both theoretically and experimentally [84]. However, we know that this process contains two stages: nucleation and growth of ice crystals. To get deeper into this process, we compared temperature profiles of the freezing process of the studied droplets (Fig. 8).

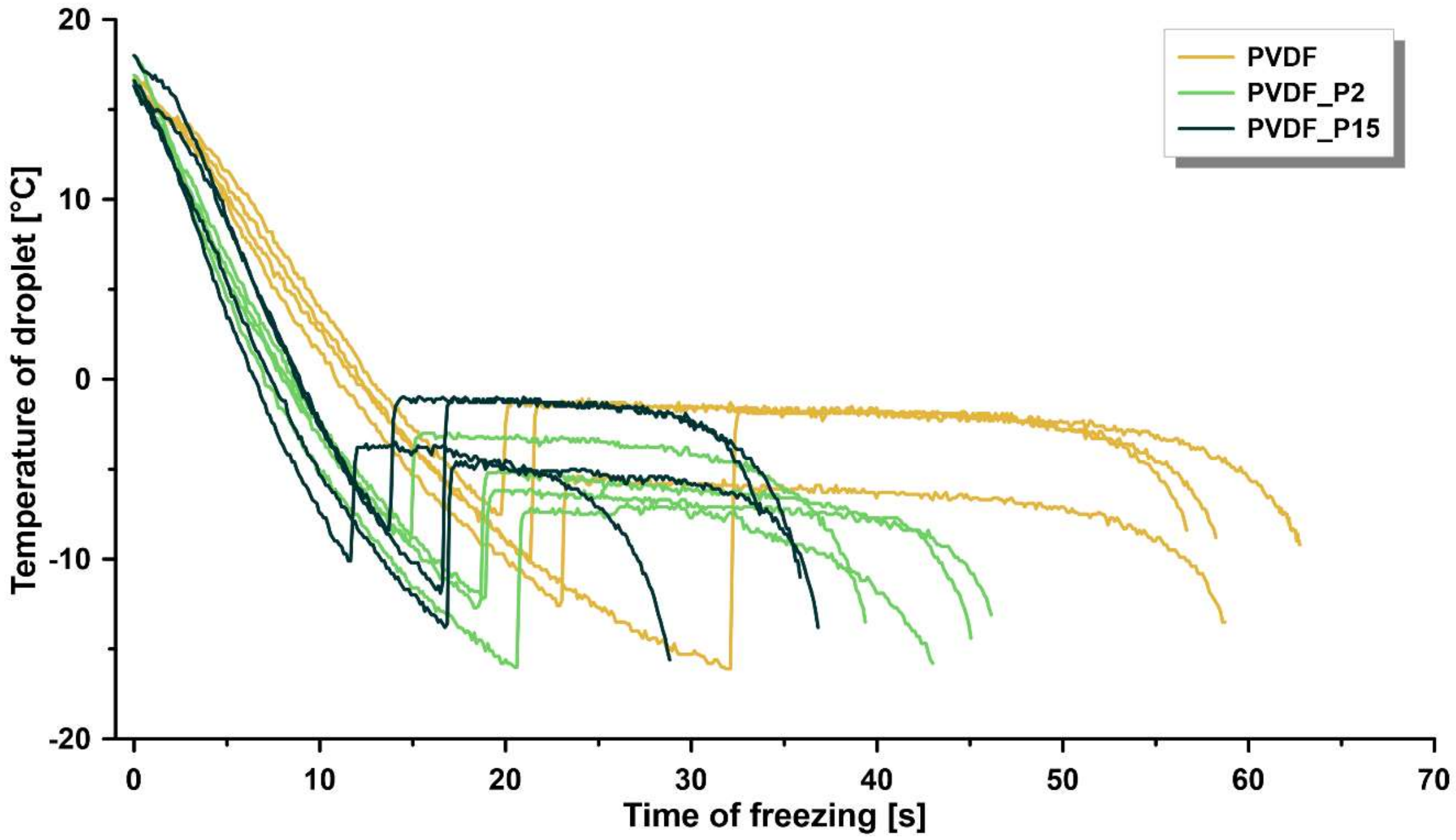

**Fig. 8.** Recorded profiles of temperature changes during the freezing process for the investigated membranes.

Obtained profiles (Fig. 8) are typical for the freezing process (see for example [84]). It is seen that the modifications performed in this study enhance the rate of droplet freezing compared to pure PVDF. From the profiles we determined two temperatures characterizing the process of recalescence, namely crystallization temperature and phase change temperature (see Fig.1 in [84]). We found a new and interesting relationship (Fig. 9), namely LEP, although not correlated with recalescence time, is linearly related to the phase change temperature.

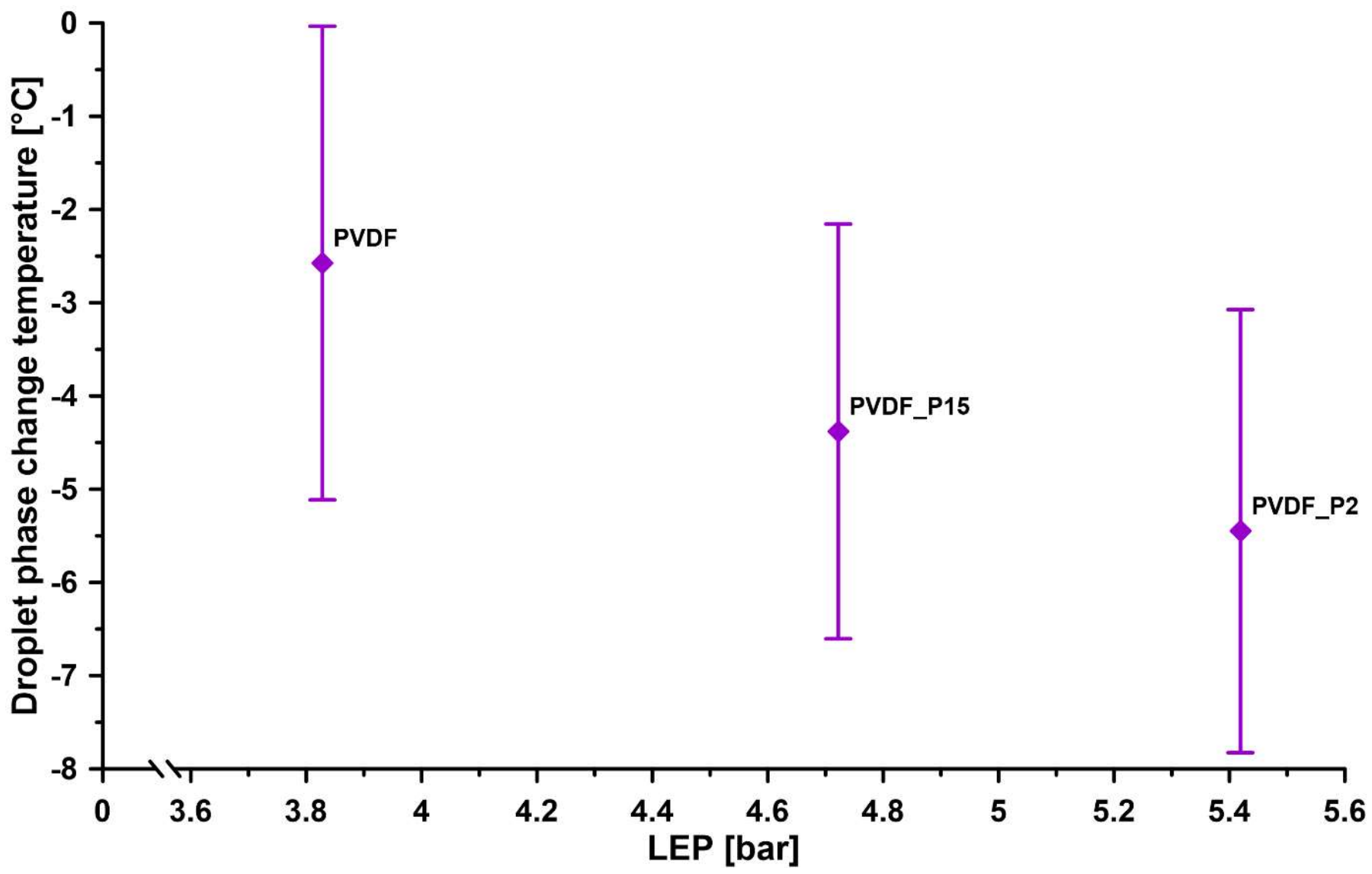

**Fig. 9**. The correlation between LEP and the droplet phase change temperature.

The latter value is more negative for larger LEP values. The existence of this correlation is an important contribution to the recalescence mechanism, which is still incomplete. As mentioned above, the physical modification leads to a simultaneous reduction of pore size, and WCA values have an opposite influence on LEP (see the Cantor- Laplace equation – Eq. S6). Therefore, the presence of $NH_2$ groups and a reduction in pore size decreases the phase change temperature compared to this recorded for pure water. This decrease is the largest (-5.45°C) for the PVDF_P2 membrane (*i.e.,* the membrane containing 2% of 1-adamantanamine). Thus, knowing LEP for studied systems, one can easily determine not only the phase change temperature during freezing but also the number of $NH_2$ groups that should be introduced to the PVDF membrane, leading (for the same Rmax) to the required phase change temperature. The mechanism of this reduction is unknown; however, it is well proved that confinement influences the structure of formed ice [85] and, in this way, freezing temperature.

Summing up, the results of this work as well as a previous study [57], show that, generally, LEP value is important in the mechanism of recalescence. For the hydrophobic systems (as studied by us previously [57]), the rise in LEP enhances the recalescence time, which is crucial for anti-freezing properties. In contrast, the introduction of polar surface centers, which also affects maximum pore size, leads to lower phase change temperature since interaction with polar surface sites enhances the heterogeneous ice nucleation process [82] by lowering interfacial free energy [86]. Since (in contrast to a previous report [57]), we observe the quasi-linear relation between time to recalescence and the freezing stage duration and concentration of polar sites, this means that the addition of hydrophilic centers promotes and speeds up the freezing process.

Recently [57], new relations were proposed between selected mechanical properties and some freezing parameters. The results of this study show that time to recalescence decreases with the rise in Young modulus, elongation at break, and toughness modulus (Fig. 10).

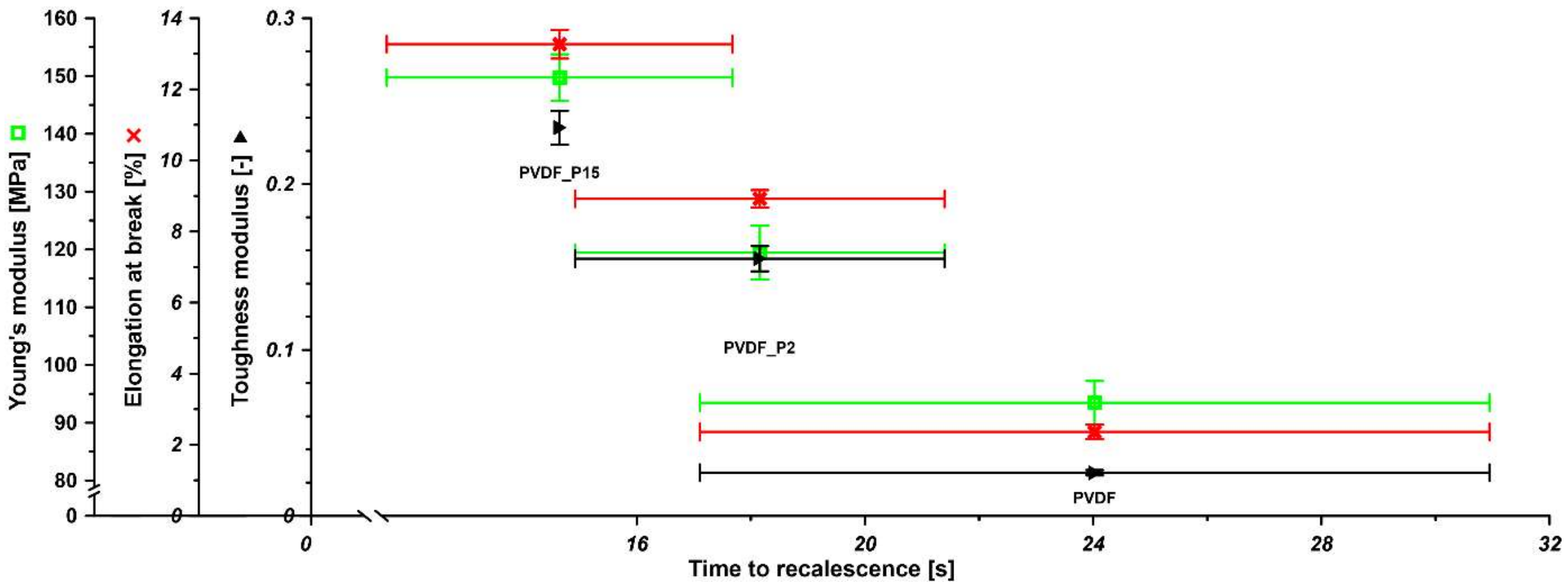

**Fig. 10**. Time to recalescence vs. Young modulus (green points), elongation at break (red points), and toughness modulus (black points).

As previously, we observe a correlation between freezing stage duration and Young and toughness modulus (Fig. 11). The observed correlation can be due to the realationship between mechanical properties and thermal conductivity of the studied membranes. This subject will be studied by us in future and the results will be reported.

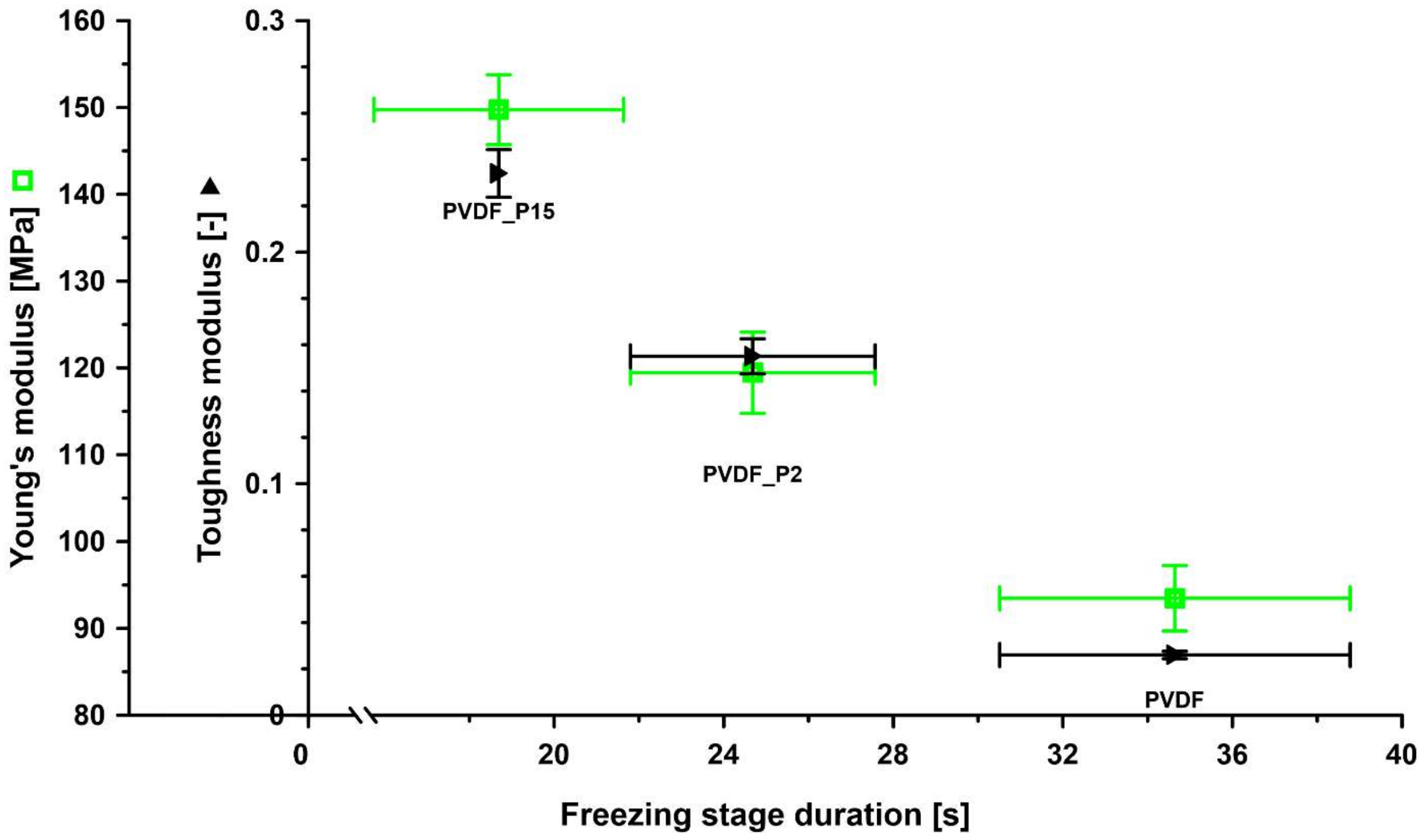

**Fig. 11**. Freezing stage duration vs. Young (green points) and toughness modulus (black points).

### 5.5 Membrane application in separation

To test the newly generated materials and check their transport and separation properties, we have chosen two different applications to perform: desalination and VOC removal. According to the established data presented in section 5.1, the membranes met the requirements of porous structure, hydrophobicity, and high resistance to wetting, which are mandatory for membrane distillation materials.

***Transport properties***

During the transport characterization, activation energy (Eq. S10), K - overall transport coefficient (Eq. S11, S12), LEP (Eq. S6), and ($p_i$/l) permeance (Eq. S14) were determined.

**Table 3**. Water transport features across the investigated membranes.

| Parameter | Membrane sample | | | | |
|---|---|---|---|---|---|
| | PVDF | PVDF_P2 | PVDF_P15 | PVDF_Ch01 | PVDF_Ch02 |
| **$\Delta p = 62.3 \pm 3.9$ mbar** | | | | | |
| **$J_{H2O}$ [kg m$^{-2}$ s$^{-1}$]** | 2.21 ± 0.07 | 3.08 ± 0.09 | 3.5 ± 0.11 | 5.30 ± 0.16 | 8.01 ± 0.24 |
| **Permeance coefficient of water at $T_f$ [kg h$^{-1}$ m$^{-2}$ bar$^{-1}$]** | 29.91 ± 1.87 | 40.60 ± 2.33 | 47.36 ± 2.92 | 71.72 ± 3.12 | 108.25 ± 3.33 |
| **K [kg m$^{-2}$ s$^{-1}$ Pa$^{-1}$]** | 0.83 ± 0.02 | 1.13 ± 0.02 | 1.32 ± 0.02 | 1.99 ± 0.03 | 3.01 ± 0.03 |
| **$\Delta p = 123.2 \pm 7.8$ mbar** | | | | | |
| **$J_{H2O}$ [kg m$^{-2}$ s$^{-1}$]** | 4.08 ± 0.12 | 5.20 ± 0.16 | 5.71 ± 0.17 | 8.43 ± 0.25 | 12.23 ± 0.29 |
| **Permeance coefficient of water at $T_f$ [kg h$^{-1}$ m$^{-2}$ bar$^{-1}$]** | 33.06 ± 1.55 | 42.17 ± 1.89 | 46.23 ± 1.84 | 68.37 ± 1.92 | 99.11 ± 2.03 |
| **K [kg m$^{-2}$ s$^{-1}$ Pa$^{-1}$]** | 0.92 ± 0.02 | 1.17 ± 0.02 | 1.28 ± 0.03 | 1.90 ± 0.02 | 2.75 ± 0.03 |
| **$\Delta p = 196.1 \pm 15.1$ mbar** | | | | | |
| **$J_{H2O}$ [kg m$^{-2}$ s$^{-1}$]** | 7.65 ± 0.22 | 8.04 ± 0.21 | 9.33 ± 0.28 | 13.22 ± 0.38 | 17.50 ± 0.45 |
| **Permeance coefficient of water at $T_f$ [kg h$^{-1}$ m$^{-2}$ bar$^{-1}$]** | 38.35 ± 2.37 | 40.10 ± 2.02 | 46.77 ± 2.55 | 66.27 ± 2.65 | 87.72 ± 3.13 |
| **K [kg m$^{-2}$ s$^{-1}$ Pa$^{-1}$]** | 1.07 ± 0.02 | 1.11 ± 0.02 | 1.30 ± 0.02 | 1.84 ± 0.03 | 2.44 ± 0.03 |
| **$\Delta p = 314.2 \pm 24.6$ mbar** | | | | | |
| **$J_{H2O}$ [kg m$^{-2}$ s$^{-1}$]** | 12.02 ± 0.34 | 12.78 ± 0.36 | 13.67 ± 0.40 | 17.88 ± 0.47 | 25.3 3± 0.67 |
| **Permeance coefficient of water at $T_f$ [kg h$^{-1}$ m$^{-2}$ bar$^{-1}$]** | 38.46 ± 2.66 | 40.96 ± 2.98 | 43.81 ± 3.03 | 57.31 ± 2.99 | 81.19 ± 3.12 |
| **K [kg m$^{-2}$ s$^{-1}$ Pa$^{-1}$]** | 1.07 ± 0.02 | 1.14 ± 0.02 | 1.22 ± 0.02 | 1.59 ± 0.03 | 2.26 ± 0.03 |

| Δp = 475.6 ± 27.6 mbar | | | | | |
|---|---|---|---|---|---|
| **$J_{H2O}$ [kg $m^{-2}$ $s^{-1}$]** | 16.23 ± 0.45 | 19.70 ± 0.54 | 21.02 ± 0.58 | 24.56 ± 0.66 | 33.52 ± 0.78 |
| **Permeance coefficient of water at $T_f$ [kg $h^{-1}$ $m^{-2}$ $bar^{-1}$]** | 34.23 ± 2.89 | 41.54 ± 3.04 | 44.29 ± 3.12 | 51.79 ± 2.99 | 70.70 ± 3.22 |
| **K [kg $m^{-2}$ $s^{-1}$ $Pa^{-1}$]** | 0.95 ± 0.02 | 1.15 ± 0.02 | 1.23 ± 0.02 | 1.44 ± 0.03 | 1.96 ± 0.03 |

The motivation was to generate appropriate hydrophobic separation materials with long-lasting high wetting resistance and transport features to reduce the problem of hampering the mainstream implementation of MD [4]. In Fig. 12, it is visible that physically and chemically modified membranes possessed improved transport features referred to the pristine PVDF material. However, the chemical modification has been more beneficial and shows higher improvement in the water transport properties (flux, permeance, and overall transport coefficient), *i.e.*, for the PVDF_Ch01 the enhancement at the lowest and the highest driving force was equal to 2.4 and 3.6 times, respectively. For the membrane PVDF_Ch02 these changes were equal to 1.6 and 2.0 times for the driving forces of 62.3 ± 3.9 mbar and 475.6 ± 27.6 mbar, respectively (Fig. 12 and Table 3). Such profound improvement was a consequence of the attachment of hydrophilic modifiers that promote the transport of water by working as a "sponge", as well as generating more rough surfaces on the membrane surface. It is essential that not only the significant transport upgrading was achieved but also the wetting resistance under high hydrostatic pressure (>11.4 bar) without compromising vapor transport was ensured (Table 1). On the other hand, the improvement for membranes filled with 1-adamantanamine in the PVDF-matrix was slightly less significant. The enhancement of the transport features was in the 1.2 – 1.6 time range, depending on the driving forces (Table 3). The important influence of the amine moieties inside and outside the membrane material has been confirmed. A similar finding has been presented by Kyoungjin An and co-workers [39], who demonstrated that the incorporation of CNT-modifiers led to the formation of hierarchical structures with both nano and micro-scale roughness. These structures minimized boundary layer formation, enhancing the driving force in CNT-modified membranes. Moreover, the Roy et al. [87] presented that the addition of CNT with the carboxylic functional group into the polymeric matrix alters the hydrophilic–hydrophobic interaction between water moisture and the membrane surface and offers a supplementary pathway for improved transport of water vapor up to 51.5% when compared with pristine materials. Additionally, Chang et al. [88] presented that increasing roughness can improve the transport features in the MD process *via* 3D printing turbulence promoters introduction to the membranes. Nevertheless, all the membranes, possessed high stability in the course of the long-lasting MD (30 days) (Fig. 13). The lack of flux reduction confirmed the stability of the hybrid membranes with 1-adamantanamine modifier. The modification also positively affected the wettability resistance (positive LEP values, higher contact angle). Moreover, manufactured membranes performed well when compared to data from the literature (Table 4).

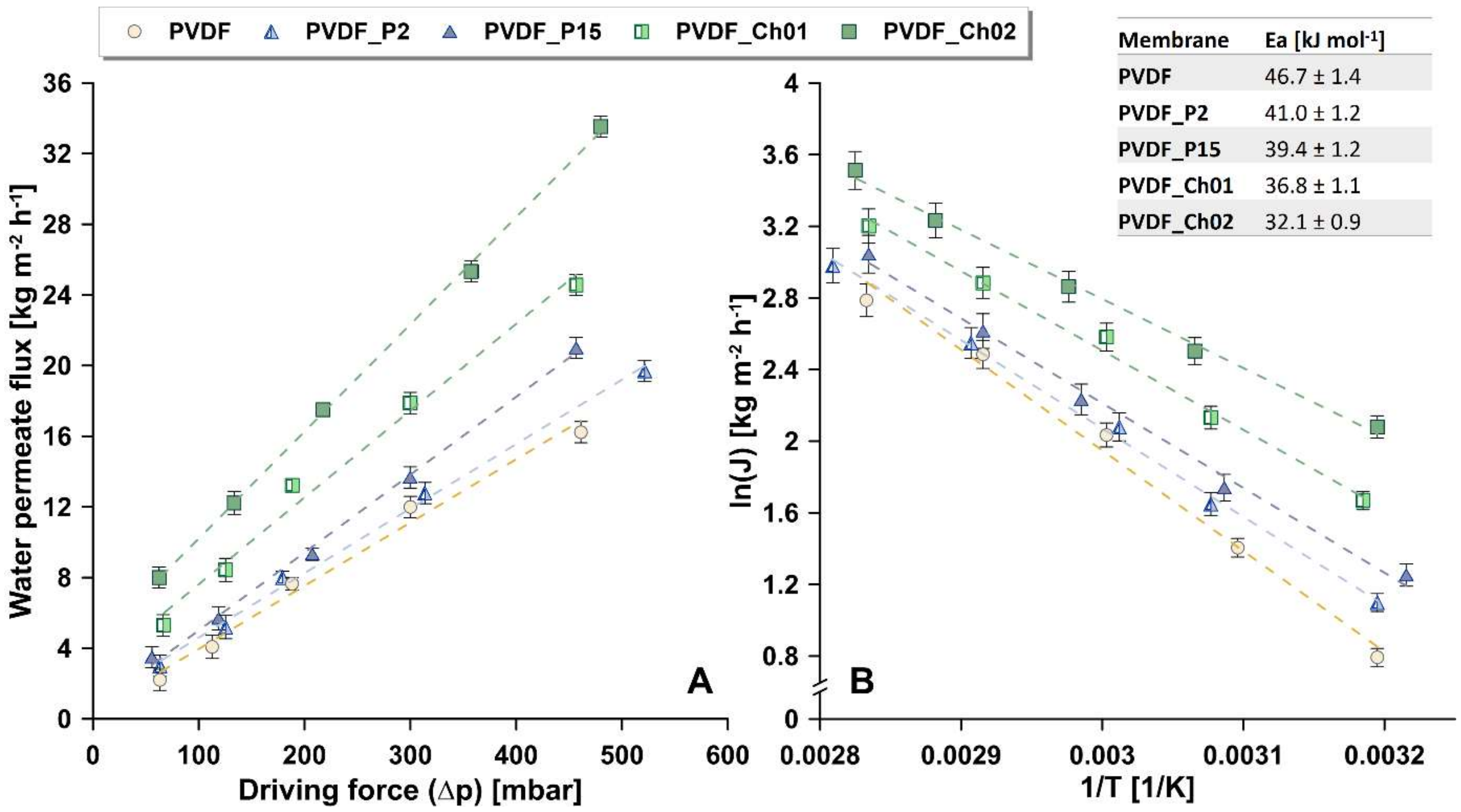

**Fig. 12.** Water transport as a function of driving force (A) and the activation energy of the membranes (B).

The concept of activation energy (Ea) allowed to gain valuable insights into the dynamics of water transport, as showcased in Fig. 12. Notably, the PVDF_Ch02 possessed the lowest Ea value of 32.1 ± 0.9 kJ $mol^{-1}$, signifying a remarkably efficient transport process that necessitates minimal energy input. However, the pristine PVDF membrane depicted the most constrained water transport, with an elevated Ea value of 46.7 ± 1.4 kJ $mol^{-1}$, alongside the lowest values of permeance coefficient and mass transfer coefficient K (Table 3). It is evident that the material composition, morphology, pore size, and tortuosity play pivotal roles in influencing the K factor. Despite the influence of temperature and pressure, the K parameter consistently maintains its value. These findings underscore the notable potential of PVDF membranes modified with 1-adamantanamine to thrive under specific driving forces, making a compelling case for their widespread application.

**Table 4.** AGMD - water transport and desalination, and material characterization - comparison with literature.

| Membrane | Modification | Flux [L m$^{-2}$h$^{-1}$] | $R_{NaCl}$ [%] | Ra [nm] | Pore size [µm] | Temp. | LEP [bar] | CA [deg] | Ref. |
|---|---|---|---|---|---|---|---|---|---|
| **Feed solution: 3.5 wt% (0.61M) NaCl** | | | | | | | | | |
| **NF-PES-8 electrospinning** | Silane nanofilament | 16.15 | 99.95 | 348 | 8 | F: 60°C<br>C: 20°C | 6.5 | 160 | [4] |
| **PTFE-0.2 electrospinning** | - | 10.65 | 99.95 | 128 | 0.2 | F: 60°C<br>C: 20°C | 6.5 | 120 | [4] |
| **PVDF/N6 dual-layer nanofiber membrane (hydrophobic surface, hydrophilic support)** | Support nylon-6 | 15.5 | 99.2 | - | 0.18 | F: 60°C<br>C: 20°C | 0.18 | 126.3 | [89] |
| **PVDF+7wt%PVP /graphene** | No Graphene<br>Graphene 0.5 wt.%<br>Graphene 1.0 wt.% | 3.54<br>5.56<br>11.3 | 99.8<br>98.0<br>99.9 | 34.2<br>94.6<br>136 | 0.182<br>0.205<br>0.164 | F: 80°C<br>C: 20°C | -1.1<br>0.12<br>0.31 | 82<br>91<br>92 | [90] |
| **PVDF-5wt.% AC** | 5 wt.% activated carbon | 0.2 | > 95 | - | 0.13 | F: 80°C<br>C: 20°C | 0.39 | 92 | [91] |
| **PVDF-PMMA** | 8 wt.% PMMA | 3.2 | 99 | 143.2 | 0.85 | F: 80°C<br>C: 20°C | 0.52 | 108 | [92] |
| **PVDF-$SiO_2$** | 5 wt.% $SiO_2$ | 3.1 | > 99 | - | 1.1 | F: 70°C<br>C: 20°C | 0.98 | 138 | [93] |
| **FGO-4/PVDF electrospinning** | 4 wt.% fluorinated graphene oxide | 21.4 | 98 | 627 | - | F: 80°C<br>C: 20°C | 0.30 | 148.3 | [94] |
| **Feed solution: 7.0 wt% (1.22 M) NaCl** | | | | | | | | | |
| **PVDF-21-Fluorosilane electrospinning** | 1 wt.% 21-Fluorosilane | 19.1 | > 99 | 65 | 0.30 | F: 70°C<br>C: 20°C | 4.12 | 151 | [95] |
| **PVDF texturing with sandpaper** | soft lithography technique for PDMS template | 12.0 | > 99 | 1500 | 0.28 | F: 70°C<br>C: 20°C | 1.35 | 116 | [96] |
| | | | | | | | | | |
| **PVDF-PLA** | Polylactic acid | 2.1 | > 99 | 570 | 1.79 | F: 65°C<br>C: 20°C | 0.49 | 128 | [97] |
| **Feed solution: 0.5M NaCl** | | | | | | | | | |
| **PVDF-SWCNH** | Coated with SWCNH (single walled carbon nano horns) | 10.2 | > 99 | 1200 | 0.30 | F: 57°C C: 8°C | 2.20 | 162 | [98] |
| **PVDF-CS-S** | Pristine chitosan connected chemically | 11.3 | > 99 | 386 | 0.14 | F: 60°C C: 10°C | 9.46 | 157.4 | [59] |
| **PVDF-CS-Q** | Partially hydrophobized chitosan connected chemically | 10.3 | > 99 | 440 | 0.17 | | 8.30 | 168.2 | |
| **PVDF – VIPS** | - | 6.38 | > 99 | 417 | 0.226 | F: 60°C C: 7°C | 3.82 | 143.2 | This work |
| **PVDF_P2** | 2 wt.% ADAM to PVDF matrix | 6.15 | > 99 | 574 | 0.198 | | 5.42 | 141.9 | This work |
| **PVDF_P15** | 15 wt.% ADAM to PVDF matrix | 7.40 | > 99 | 539 | 0.132 | | 4.72 | 131.7 | This work |
| **PVDF_Ch01** | 0.1 wt.% ADAM chemically grafted on the surface | 11.40 | > 99 | 693 | 0.152 | | 8.11 | 148.9 | This work |
| **PVDF_Ch02** | 0.2 wt.% ADAM chemically grafted on the surface | 15.09 | > 99 | 756 | 0.121 | | 11.45 | 164.2 | This work |

F – feed temperature; C – cooling temperature
AC – activated carbon
ADAM - 1-adamantanamine
21-Fluorosilan - Trichloro(3,3,4,4,5,5,6,6,7,7,8,8,9,9,10,10,11,11,12,12,12-henicosafluorododecyl)silane

***Desalination process***

The thoroughly characterized membranes were utilized for the desalination of a 0.5 M NaCl solution, which closely mimicked seawater salinity [99]. Similar to the pure water fluxes, the membranes during the desalination process possessed a stable flux over the entire experiment (Fig. 13). As in the case of the prior tests, the measurements were done at various driving forces to thoroughly test the membranes, ensuring a comprehensive analysis of their performance. The disparity in permeate flux between salty water and pure water can be attributed to the principle of MD, where only solvent vapors can permeate through the hydrophobic porous structure of the membrane (Fig. 13). This reduction in transport is governed by Raoult's law. In membrane distillation, the transport of solvent vapors is directly linked to the difference in water vapor pressure between the feed and permeate sides. It is important to note that, regardless of the feed type, an enhancement in transport features is observed with an increase in driving force (Fig. 13).

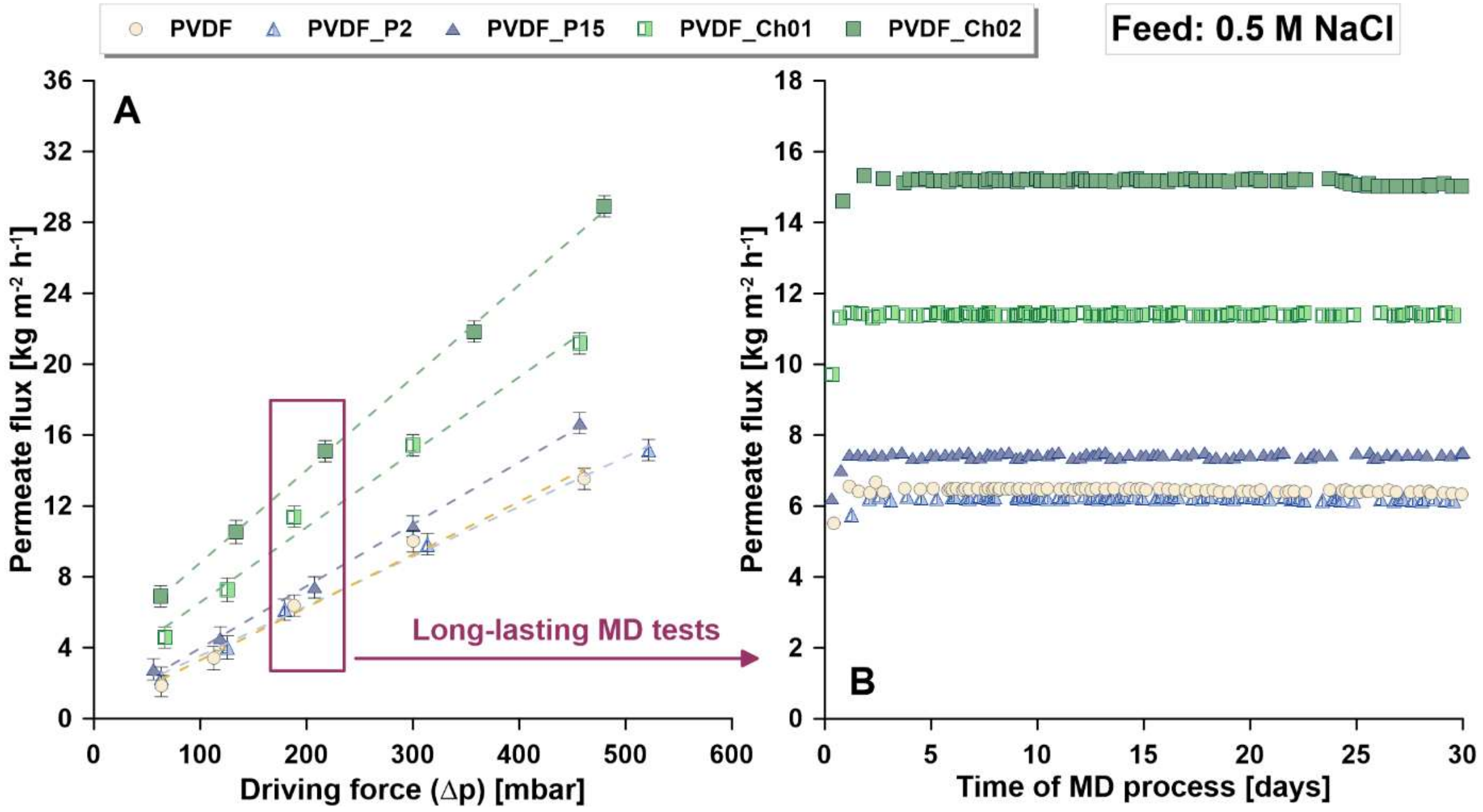

**Fig. 13.** Transport properties of the synthesized membranes. Driving forces range 60 – 480 mbar (A). Long-lasting MD test – 30 days, tests at driving force ca. 200 mbar (B). Feed 0.5 M NaCl.

The salt rejection coefficient was measured to examine the efficiency of desalination. The gained data showed that all the generated membranes have very high $R_{NaCl}$ parameters, which varied within 98.7–99.9%. This result verifies no wetting effect during the MD process. In order to verify this statement, the t-3 ing force was tested using a saltwater solution over an extended duration. 200 mbar (Fig. 13 B). The obtained values of the separation and transport features confirm the development of highly efficient membranes. To contextualize these findings, pertinent literature was compiled in Table 4. During the comparison, membranes with the same geometry and tested at the same driving force, showed lower transport abilities (Table 4).

### *Hazardous VOCs removal*

Two model solutions were selected, *i.e.*, $H_2O$-EtOH and $H_2O$-MTBE. REACH considers both as hazardous VOCs [100] (Table S7). According to the data presented above, the resistance to wetting by water has been justified. Nevertheless, the potential risk of wetting or swelling by organic components must be determined in removing VOCs, which was accomplished by implementing the HSP approach. To evaluate the interaction between the membrane material and the separation system, calculations were performed for pure VOCs, specifically ethanol and MTBE (Table S8), as well as for feed solutions containing up to 5 wt.% of the organic component (Table S9). In the case of MTBE affinity, the modified materials exhibited the strongest interaction that was confirmed by a noticeably shorter distance in the Hansen sphere (Ra, Δi.membrane parameter) (Table S8 and S9). The reduction in the Ra parameter between the pristine membrane and PVDF_Ch02 was as follows: Ra_PVDF = 9.93 $MPa^{1/2}$ and Ra_PVDF_Ch02 = 6.08 $MPa^{1/2}$. Such a tendency was related to the nonpolar character and lower dielectric constant (ε) of the MTBE (Table S7), and the enhancement in the affinity with more hydrophobic materials was noticed [101]. On the other hand, affinity to ethanol and feed aqueous solution was slightly higher for less hydrophobic materials, *i.e.*, membranes treated *via* physical way and PVDF_Ch01. This phenomenon was related to the higher polarity of the mixtures and much higher value of ε (ε for $H_2O$-EtOH(5 wt.%) = 75.7, ε for $H_2O$-EtOH(5 wt.%) = 74.6) (Table S7). The stability was evaluated by the RED values (Eq. S9) and was higher than 1 for all investigated membranes in contact with feed solutions (Table S9). The RED values for water-EtOH mixtures were in the range of 3.89 – 4.24 and water-MTBE in the range of 3.80 – 4.15. High resistance to potential wetting of the material with the tested liquids was confirmed. Moreover, a slightly better affinity was visible for the MTBE-based mixture owing to lower RED values (Table S9). Based on established data, all membranes are proven to be highly suitable for use in the MD process, guaranteeing zero risk of wetting or destruction.

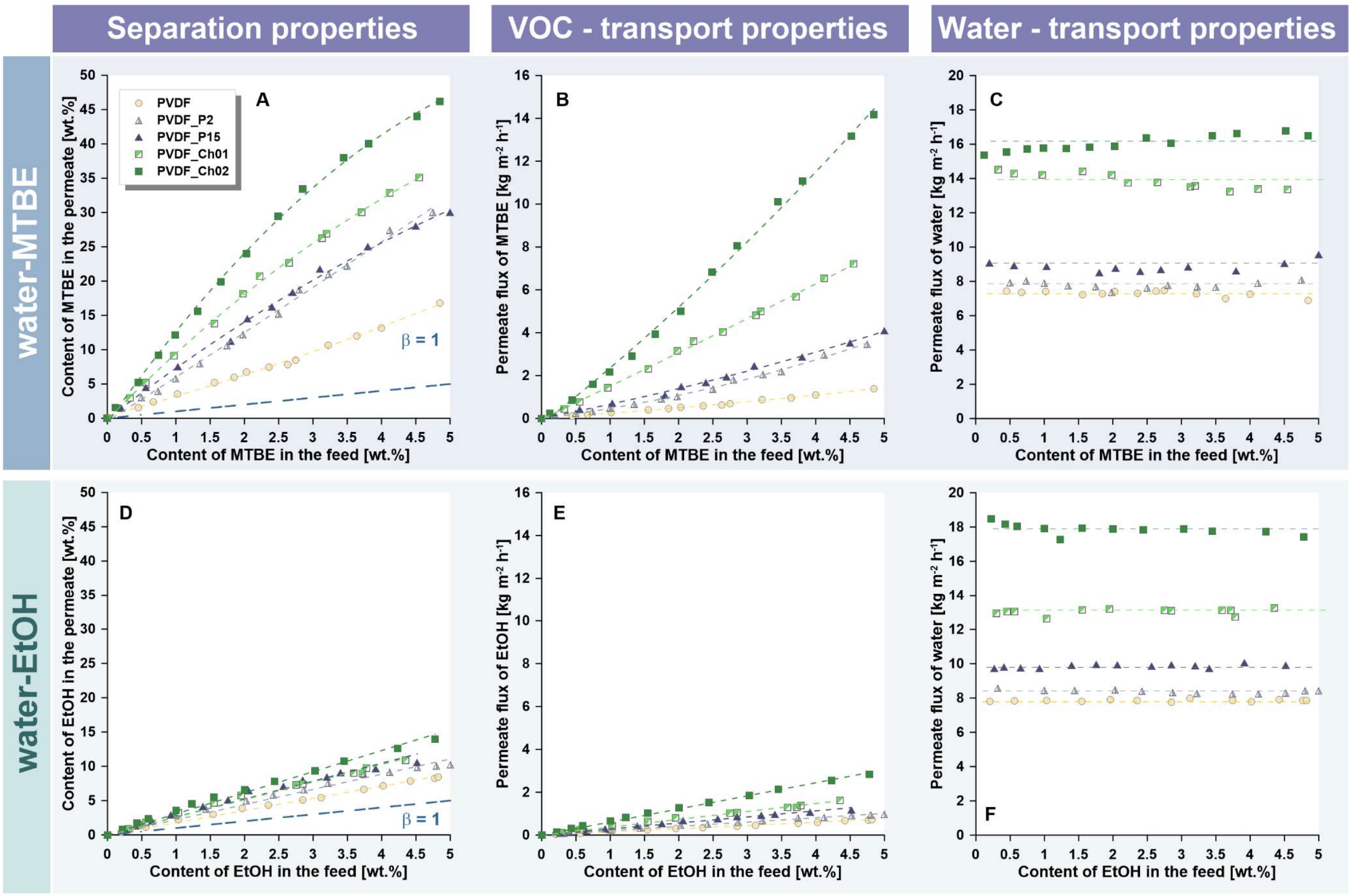

**Fig. 14**. Separation McCabe-Thiele diagrams (A, D) and transport properties of the investigated membranes (B, E – organic component transport, C, F – water transport).

All membranes showed high selectivity to the organic constituents, obviously evidenced by a separation factor β above 1 (Fig. 14A, 14D). The expression of the parameters describing membrane performance in MD process have been well modeled by Eqs. S15-S18. Moreover, HSP analysis in general (Table S9) indicated a clear organic solvent preference. The experimental results from the MD process align closely with the HSP calculation predictions. Notably, the modification led to a significant enhancement in both transport and separation performance when compared to unmodified PVDF (Fig. 14).

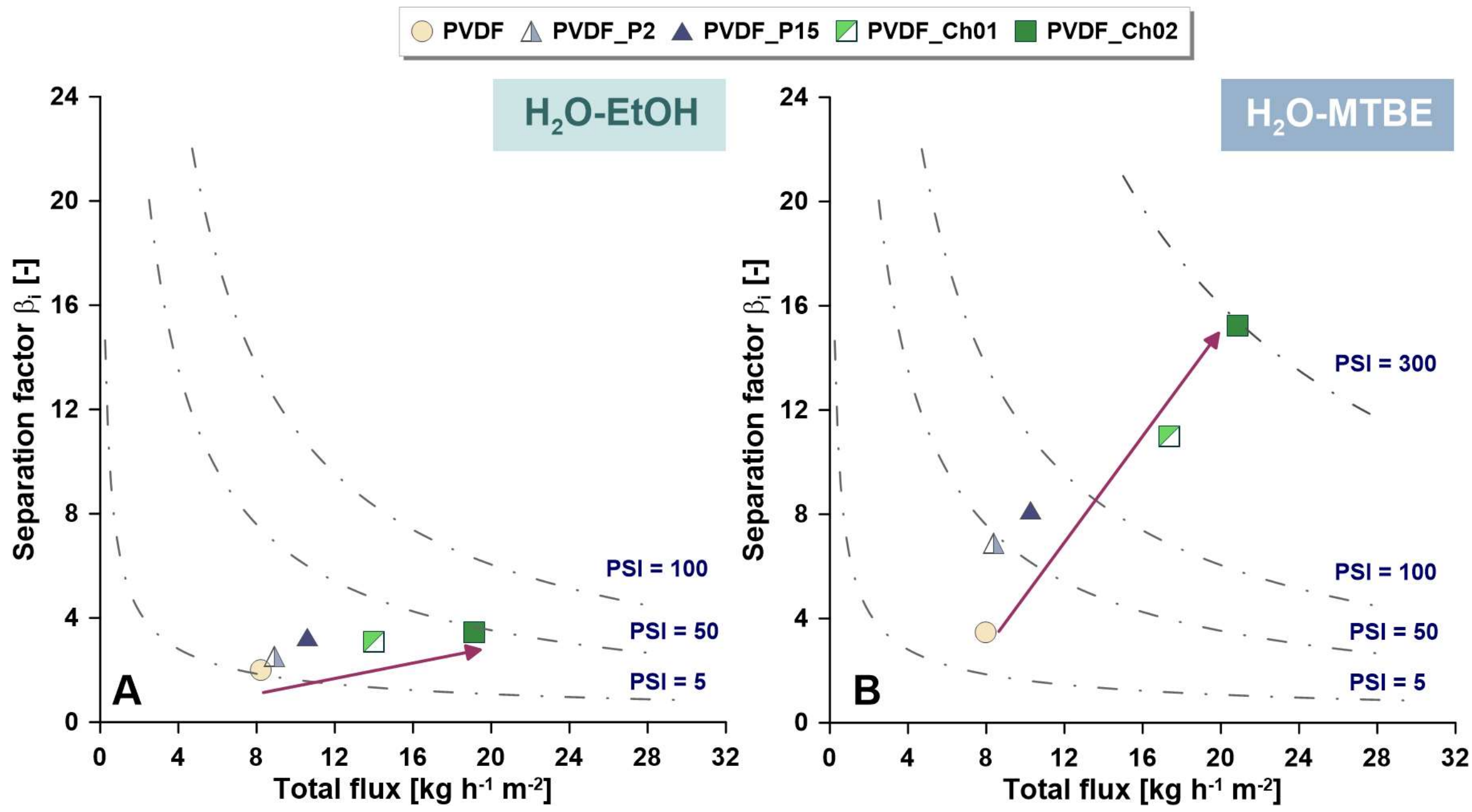

**Fig. 15**. Membrane separation performance (2.0 wt.% of EtOH or MTBE in feed). PSI unit [kg m$^{-2}$ h$^{-1}$].

The best performance was observed for the PVDF_Ch02 for both separated systems, *i.e.*, $H_2O$-EtOH and $H_2O$-MTBE (Fig. 14, 15). Transport and separation features were enhanced, which can be elucidated by creating the extra layer of the modifier (Fig. 2) that promotes water transport. Moreover, there is an amendment in surface roughness and affinity between 1-adamantanamine and water (Table S4, S5). A slightly smaller improvement for PVDF_P2 and P15 is caused by the location of the modifier inside the polymeric matrix. PVDF_Ch02 membrane possesses the PSI factor of 297 kg $m^{-2}$ $h^{-1}$ for MTBE removal and ca. 47 kg $m^{-2}$ $h^{-1}$ for EtOH removal from water (Fig. 14, 15). Overall, the $H_2O$-MTBE system shows better separation efficiency compared to $H_2O$-EtOH, which was consistent with the HSP analysis. However, this phenomenon demands further examination due to the polarity of the VOCs: $\varepsilon$MTBE = 2.6 < $\varepsilon$EtOH = 24.5 << $\varepsilon H_2O$ = 78.4 [76]. The significantly lower $\varepsilon$ for MTBE is a key factor contributing to its superior separation efficiency. To illustrate these differences more comprehensively, Fig. 14 A and D present McCabe-Thiele diagrams. The established data for polymeric membranes are much higher than those available in the literature (Table 5).

**Table 5.** MD – VOCs removal efficiency - comparison with literature.

| Membrane | MD mode | Flux [kg m$^{-2}$h$^{-1}$] | β [-] | PSI [kg m$^{-2}$h$^{-1}$] | Temp. | VOC [wt.%] | Ref. |
|---|---|---|---|---|---|---|---|
| | | | **Ethanol removal** | | | | |
| **PVDF 0.45 µm** | AGMD, 4 mm gap | 11.1 | 2.7 | 18.7 | F: 60°C<br>C: 20°C | 2.0 | [102] |
| **Polypropylene (PP)** | DCMD | 4.4 | 3.32 | 10.21 | F: 60°C<br>C: 20°C | 2.0 | [103] |
| **GO membranes** | VMD | 2.5 | 3.02 | 5.0 | F: 60°C<br>C: 20°C | 2.0 | [104] |
| **PP V8/2 HF 0.2 µm** | DCMD | 12.2 | 7.0 | 8.4 | F: 60°C<br>C: 20°C | 4.0 | [105] |
| **PP S6/2 0.2 µm** | DCMD | 14.6 | 7.6 | 11.5 | F: 60°C<br>C: 20°C | 4.0 | [105] |
| **PTFE** | VMD | 24.14 | 8.6 | 183.4 | F: 40°C<br>C:20°C | 2.0 | [106] |
| **PTFE 0.2 µm** | AGMD, 3.12 mm gap | 4.1 | 3.7 | 11.07 | F: 55°C<br>C: 20°C | 2.0 | [107] |
| | | | **MTBE removal** | | | | |
| **PTFE 0.2 µm** | SGMD | 0.75 | 31 | 22.5 | F: 40°C | 1.5 | [108] |
| **CNIM-f** | SGMD | 0.82 | 36 | 28.7 | F: 40°C | 1.5 | [108] |
| **GOIM** | SGMD | 1.40 | 46 | 63 | F: 40°C | 1.5 | [108] |
| **PTFE 0.5 µm** | VMD | 0.85 | 6 | 4.25 | F: 40°C<br>C: 20°C | 2.0 | [109] |
| **$ZrO_2$- n-octyltrichlorosilane** | VMD | 0.6 | 12 | 6.6 | F: 60°C<br>C: 20°C | 2.0 | [77] |
| **$ZrO_2$-1H,1H,2H,2H-perfluorooctyltrichlorosilane** | VMD | 1.2 | 15 | 16.8 | F: 60°C<br>C: 20°C | 2.0 | [77] |
| **$TiO_2$- n-octyltrichlorosilane** | VMD | 2.0 | 48.2 | 93.2 | F: 60°C<br>C: 20°C | 2.0 | [19] |
| **$TiO_2$-trichloro(octadecyl)silane** | VMD | 2.1 | 38.5 | 79.4 | F: 60°C<br>C: 20°C | 2.0 | [19] |
| **$TiO_2$-1H,1H,2H,2H-perfluorooctyltrichlorosilane** | VMD | 2.0 | 56.4 | 111.1 | F: 60°C<br>C: 20°C | 2.0 | [19] |
| **$TiO_2$-activation with piranha + n-octyltrichlorosilane** | VMD | 8.08 | 94.2 | 754.5 | F: 60°C<br>C: 20°C | 2.0 | [19] |
| **$TiO_2$-activation with piranha + trichloro(octadecyl)silane** | VMD | 4.55 | 89.1 | 401.2 | F: 60°C<br>C: 20°C | 2.0 | [19] |
| **$TiO_2$-activation with piranha + 1H,1H,2H,2H-perfluorooctyltrichlorosilane** | VMD | 6.32 | 109.5 | 658.3 | F: 60°C<br>C: 20°C | 2.0 | [19] |

F - feed temperature; C - cooling temperature

CNIM-f - functionalized carbon nanotube-immobilized membranes; GOIM - graphene oxide-immobilized membranes; SGMD – sweeping gas MD

***Membrane stability and durability***

The Hansen Solubility Parameters were assessed to control the stability of the membranes, and further tests were conducted to check their durability. These tests were applied to understand the influence of the AGMD process with VOCs on the features of membrane materials. Each membrane underwent five tests, each lasting at least 50 hours. The permeate flux, WCA, and roughness were precisely checked after each test. The results revealed membrane stability (Fig. 16).

However, it is worth noting that minor variations in stability were noticed between the pristine PVDF and the modified samples. The chemically modified samples (PVDF_Ch01 and PVFD_Ch02) exhibited more excellent resistance. A significant 25% reduction in transport properties for the unmodified PVDF between the first and last tests was noticed. In contrast, the permeate flux reduction for the PVDF_Ch01 and PVDF_Ch02 materials was merely 9% and 6%, respectively. Moreover, the differences in contact angle ranged from 5% for the modified membranes to 8% for PVDF. A positive note is that the roughness parameters showed only negligible changes, remaining below 5% for all the investigated membranes. These findings confirm that, although a slight drop in transport was detected, the extended testing did not affect the grafted layer or the highly hydrophobic nature of the PVDF-based membranes. The high stability of the grafted layer in the case of PVDF_Ch02 has been additionally confirmed by the FTIR and SEM of the membranes used in the stability tests (Fig. 16 C, D). In addition, the stable WCA and roughness values unequivocally demonstrate the exceptional stability of the covalent modification.

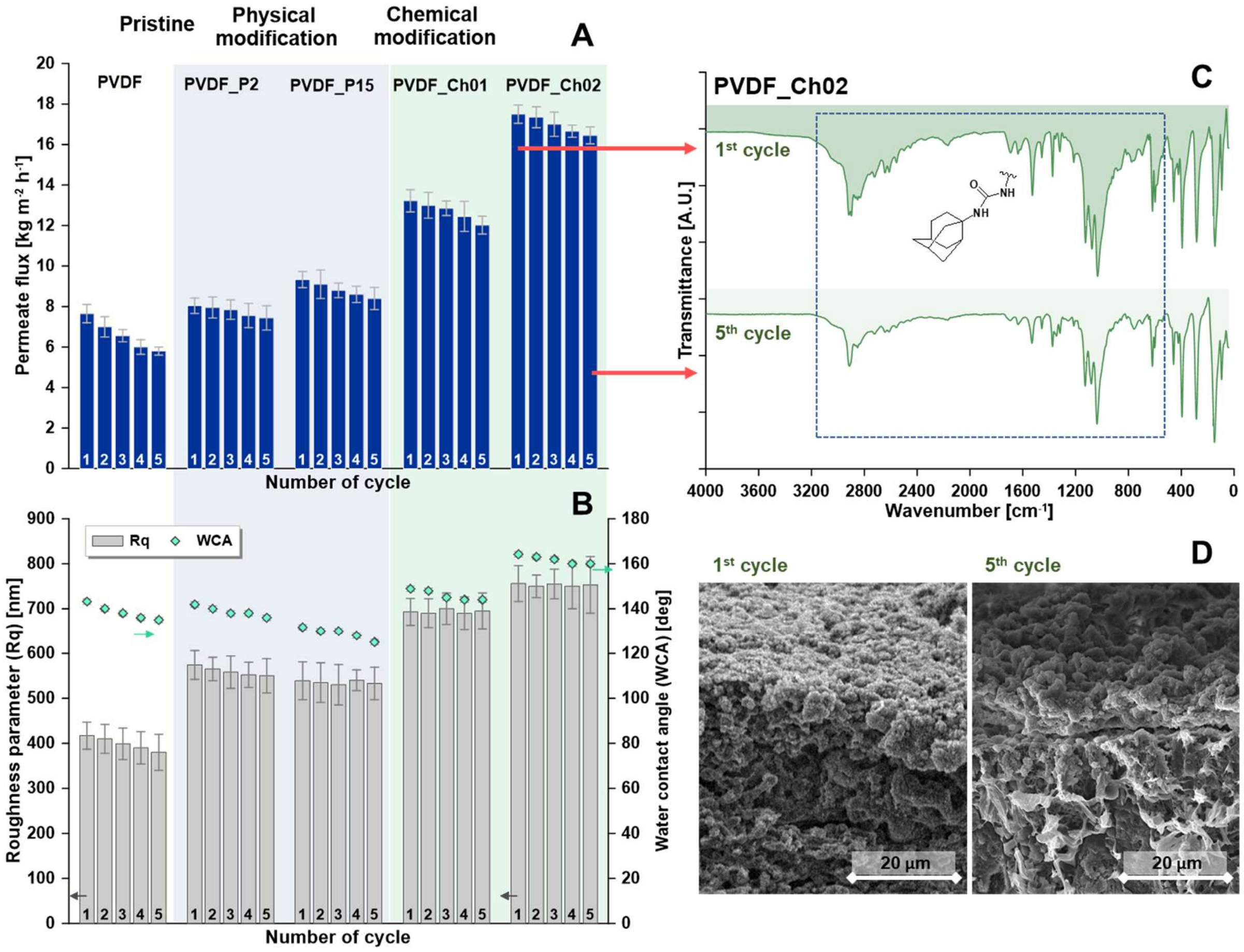

**Fig. 16.** Membrane stability– transport and material properties (A,B). Stability of the grafted layer on PVDF_Ch02 – FTIR (C) and SEM (D).

### *Assessment of bioaccumulation potentials*

When utilizing modified membrane materials in water and wastewater treatment, it is crucial to assess the potential risk of bioaccumulation [110]. The widespread use of long-chain perfluoroalkylsilanes has resulted in a concerning accumulation in wildlife, primarily through direct uptake *via* gills or diet (bioaccumulation or bioconcentration). This highlights the urgent need for action to address the environmental impact of these substances [111]. The substantial bioaccumulation potential and acute toxicity of long-chain perfluoroalkylsilanes has ignited global concern, prompting strict restrictions on their usage in various sectors [112, 113]. In contrast, short/ultrashort-chain fluorocarbon molecules are heralded for their lower acute toxicity and heightened biosafety [114]. Yet, the superior choice lies in replacing fluorinated modifiers with fluorine-free modifiers capable of conferring similar or even enhanced surface properties. The bioaccumulation potentials of chosen modifiers was assessed by examining log BAF, log BCF, and log Kow factors (Fig. 17). Notably, the bioaccumulation potential (highlighted by the red zones in Fig. 17) steeply increases with fluorinated molecules, particularly those with longer chains, as denoted by elevated log Kow values (Fig. 17A). Conversely, ultrashort-chain fluorocarbons, some naturally occurring modifiers, and the non-fluorinated silane, as well as the "green modifier" utilized in the current study exhibit non-bioaccumulative characteristics, with log BAF, log BCF, and log Kow values below 3.7, 3, or 5 (highlighted by red in Fig.17), respectively [115, 116].

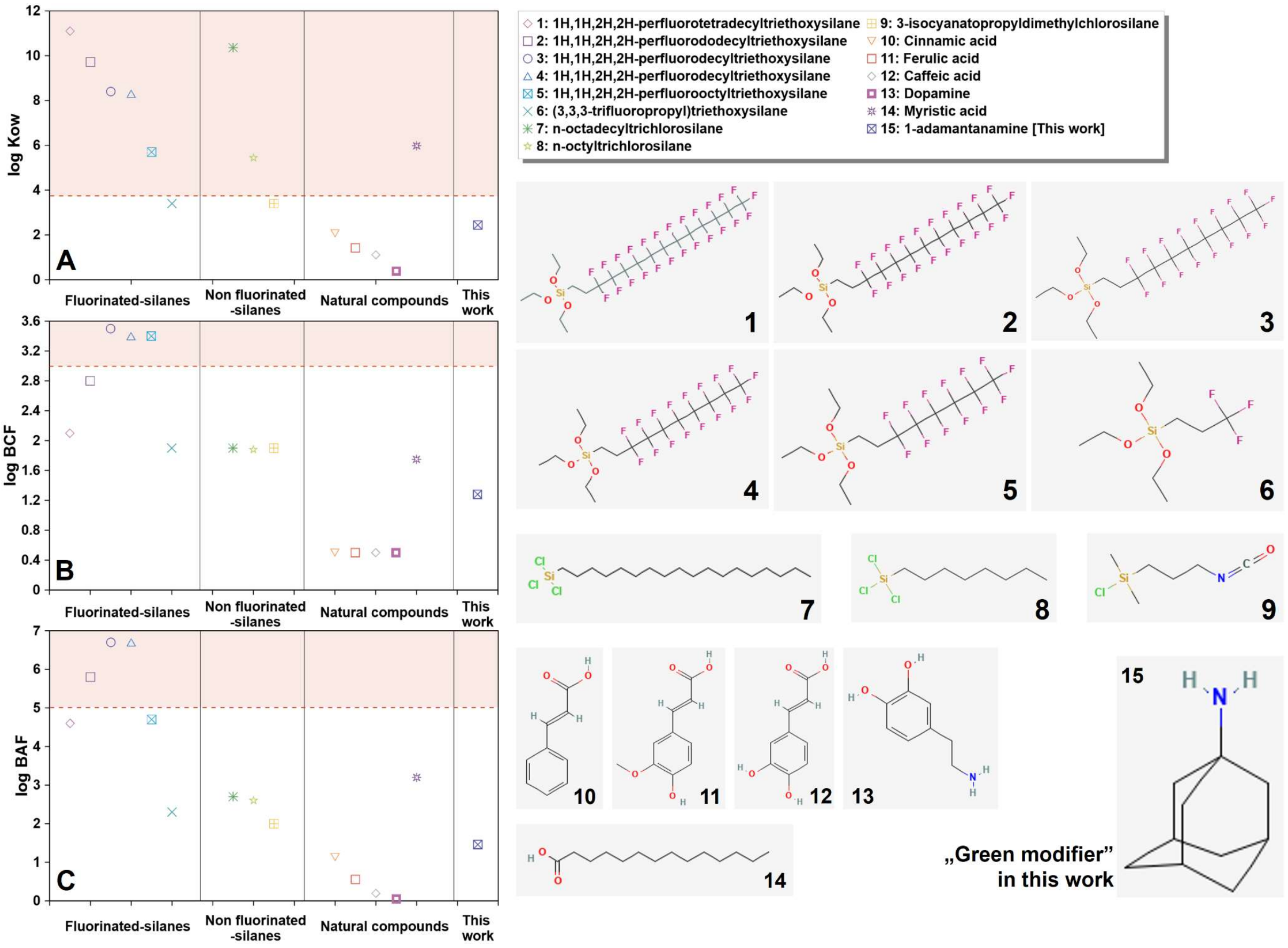

**Fig. 17.** Bioaccumulation potential with the structure of selected membrane modifiers for MD applications. (Log BCF values are calculated by a regression-based method [117]. Log BAF values are determined by the Arnot–Gobas approach (upper trophic) [116].)

## 6 Conclusions

Utilizing the principles of biomimicry, hybrid separation materials with green modifiers were developed. The addition of 1-adamantanamine, either physically or chemically, vastly improved the membranes' transport and separation capabilities in MD application by introducing surface pine-like structures. These meticulously crafted structures, obtained using a green modifier (producing eco-friendly membranes), drastically increase hydrophobicity, and show very good performance in removing hazardous VOCs such as EtOH and MTBE from water, contributing to a cleaner and safer environment. The introduction of 1-adamantanamine has, simultaneously, notably enhanced transport and separation capabilities, thus, the transport was enhanced above 3.6 times, and the separation factor beta changed from 3.48 to 15.22 for MTBE removal and from 2.0 to 3.46 for EtOH removal when comparing pristine PVDF with the membrane chemically modified with 1-adamantanamine (PVDF_Ch02). The process separation index during the MTBE removal changed from 20 kg $m^{-2}$ $h^{-1}$ (PVDF) to 297 kg $m^{-2}$ $h^{-1}$ (PVDF_Ch02). For EtOH removal, the PSI rose from 8.2 kg $m^{-2}$ $h^{-1}$ (PVDF) to 47 kg $m^{-2}$ $h^{-1}$ (PVDF_Ch02). PVDF_Ch02 was the most effective membrane. The material with the best efficiency and durability was prepared via chemical modification with the higher concentration of modifier, *i.e.*, 0.2M. All materials were highly stable and durable during the MD applications, and the implementation of new membranes in the study of droplet freezing on a cold plate allowed to draw new conclusions describing the mechanistic aspects of this process. Summing up, since 1-adamantanamine promotes transport and enhances the affinity to the VOC, ensuring excellent membrane performance at different applications of the MD process, it is a very promising eco-friendly compound for future membrane improvement.

## Acknowledgments

J.K. gratefully acknowledges the financial support from Excellence Initiative - Mobility for Employees competition as part of the "Excellence Initiative - Research University" program (77/2023/Mobilność; 90-SIDUB.6102.27.2023.MP8). The research was supported partially by 2018/29/B/ST4/00811 (Opus 15) grant from the National Science Center, Poland. A.P.T gratefully acknowledges the financial support from NCN (National Science Centre, Poland) Opus 22 project: UMO-2021/43/B/ST5/00421.